\documentclass[a4paper,11pt]{article}
\pdfoutput=1 

\usepackage{jcappub} 

\usepackage[T1]{fontenc}
\usepackage{color}

\newcommand{\uvec}[1]{\hat{\mathbf{#1}}}
\newcommand{\gELM}[0]{\gamma^{E}_{LM}}
\newcommand{\gBLM}[0]{\gamma^{B}_{LM}}
\newcommand{\GLM}[0]{{}_{\pm}\gamma_{LM}}
\newcommand{\Xlm}[1]{{}_{\pm}X_{#1}}
\newcommand{\gELMest}[0]{\widehat{\gamma}^{E}_{LM}}
\newcommand{\gBLMest}[0]{\widehat{\gamma}^{B}_{LM}}
\newcommand{\T}[0]{T(\uvec{n})}
\newcommand{\X}[0]{{}_{\pm}X(\uvec{n})}

\newcommand{\dX}[0]{{}_{\pm}\delta X(\uvec{n})}
\newcommand{\Q}[0]{Q(\uvec{n})}
\newcommand{\U}[0]{U(\uvec{n})}
\newcommand{\dist}[1]{{}_{\pm}\delta X_{#1}}
\newcommand{\gQ}[0]{\gamma^{Q}(\uvec{n})}
\newcommand{\gU}[0]{\gamma^{U}(\uvec{n})}

\newcommand{\eff}[0]{f^{L}_{l_{2}}}
\newcommand{\bobs}[1]{B^{\rm obs}_{#1}}
\newcommand{\tobs}[1]{T^{\rm obs}_{#1}}

\newcommand{\Ipm}[2]{{}_{\pm}I^{#1}_{#2}}
\newcommand{\Iplus}[2]{{}_{+}I^{#1}_{#2}}
\newcommand{\Iminus}[2]{{}_{-}I^{#1}_{#2}}

\newcommand{\Yspin}[1]{{}_{\pm 2}Y_{#1}(\uvec{n})}

\usepackage{caption}
\usepackage{subcaption}
\usepackage{graphicx}
\usepackage{dcolumn}
\usepackage{bm}
\usepackage{float}

\graphicspath{{"./figures/"}}

\title{\boldmath Blind Map Level Systematics Cleaning: A Quadratic Estimator Approach}

\author[a]{Joel Williams}

\author[a]{Nialh McCallum}
\author[a]{Aditya Rotti}
\author[a,b]{Daniel B. Thomas}
\author[a]{Richard Battye}
\author[a]{Michael L. Brown}

\affiliation[a]{
 Jodrell Bank Centre for Astrophysics, Department of Physics and Astronomy, School of Natural Sciences, University of Manchester, Manchester, M13 9PL, U.K.
}%
\affiliation[b]{School of Physics and Astronomy, Queen Mary University of London, London, E1 4NS, UK}

\emailAdd{christopher.williams-4@manchester.ac.uk} 
\date{\today}

\abstract{We present the first detailed case study using quadratic estimators (QE) to diagnose and remove systematics present in observed Cosmic Microwave Background (CMB) maps. In this work we focus on the temperature to polarization leakage. We use an iterative QE analysis to remove systematics, in analogy to de-lensing, recovering the primordial B-mode signal and the systematic maps. We introduce a new Gaussian filtering scheme crucial to stable convergence of the iterative cleaning procedure and validate with comparisons to  semi-analytical forecasts. We study the limitations of this method by examining its performance both on idealized simulations and on more realistic, non-ideal simulations, where we assume varying de-lensing efficiencies. Finally, we quantify the systematic cleaning efficiency by presenting a likelihood analysis on the tensor to scalar ratio, $r$, and demonstrate that the blind cleaning results in an un-biased measurement of $r$, reducing the systematic induced B-mode power by nearly two orders of magnitude.}

\begin{document}
\maketitle
\flushbottom
\section{Introduction}
\label{sec:introduction}

The Cosmic Microwave Background (CMB) intensity and polarization are key observables for cosmology. The frontiers of cosmology have been pushed back by progressively more sensitive measurement of the CMB delivered by Cosmic Background Explorer (\textit{COBE}) \citep{1996ApJ...464L...1B}, the Wilkinson Microwave Anisotropy Probe (\textit{WMAP}) \citep{2013ApJS..208...20B}, and \textit{Planck} \citep{2018arXiv180706205P}. Future surveys, including satellite and ground-based, will measure the CMB sky with unprecedented sensitivities, with the primordial B-mode polarization spectrum as one of the primary targets. With these ever increasing sensitivities, precision control over systematics and their removal will become increasingly important. 

Robust measurement of the primordial B-mode signals will require overcoming a number of analysis challenges, including foreground removal, removal of the weak-lensing B-mode signal and  potential contamination from instrument systematic effects. Recent studies have shown that the upcoming experiments in principle have sufficient sensitivity and frequency coverage for robust recovery of B-mode signal corresponding to $r \sim 10^{-3}$ \citep[e.g.][]{Ade:2018sbj, Remazeilles2020cMILC}. Similarly other studies have examined how well B-mode skies may be de-lensed \citep[e.g.][]{Sherwin2015}. In this work we focus our attention on systematics from instruments, specifically examining them using quadratic estimators (QE).

Common approaches to the removal of instrument systematics require complex modeling and prior knowledge of the instrument itself. An appealing aspect of a QE approach is that it in principle allows a largely agnostic approach to dealing with instrument systematics. That is, the effect of systematics on the CMB data can be modeled as a set distortions to the CMB data. Reconstructing these distortions using a QE does not require prior knowledge or modeling of the instrument sourcing the distortions. QE cleaning and reconstruction is therefore a promising complimentary approach to traditional systematic modeling techniques.
\par 
Previous works have suggested the use of QE as a method to quantify the level of systematics in CMB maps \citep[e.g][]{2010PhRvD..81f3512Y}, formulating QE for a variety of instrumental systematics. These included  gain fluctuations, differential systematics, and instrumental polarisation rotation, to list a few among a much longer list of possible instrument systematics. QEs are most commonly employed in reconstruction of the lensing potential map \citep[e.g.][]{planck2020_lensing} but have also been used to constrain cosmological birefringence \citep[e.g.][]{Gluscevic2012,Ade2015, Array2017}. QE studies frequently draw from understanding gained from lensing reconstruction. We improve upon previous studies of QEs as applied to systematic effects by considering an estimator in the full sky regime, accounting for realistic scan patterns, and testing whether certain aspects of the conventional wisdom from lensing studies apply. We consider two scenarios throughout this work: (i) a no lensing, noise, and beam free scenario which we refer to as the ``ideal case'', and (ii)~a scenario we refer to as the ``more realistic, non-ideal case'' that includes the effect of lensing on the CMB spectra, a gaussian white noise of $w^{-1}_{TT}=2.7\mu \text{K}\,\text{arcmin}$ and a full width at half maximum of $\theta_{\text{FWHM}}=30^{\prime}$. This is motivated by the effective noise and beam expected for the \emph{LiteBIRD} experiment \cite{Hazumi2019}.
%
\par
The QE approach would, in principle, leave us to deal with many systematics. In practice, to understand the most relevant ones it is useful to use rough estimates of the expected contamination sources and then deal with those that are most prominent. A potentially large source of CMB contamination is a temperature to polarization (T to P) leakage caused by a differential gain systematic. Since the CMB temperature anisotropy signal is 3-4 orders of magnitude larger than primordial B-mode signal, even a small leakage can induce large B-mode power. Therefore, this systematic may be a 
large hindrance for primordial B-mode studies. In this work we will refine aspects of analysis presented in \citep{2010PhRvD..81f3512Y}, with our detailed scrutiny limited to focusing on this systematic. 
%

\par
We reiterate that, while all tasks necessary for controlling and understanding this instrument systematic will be performed, it is essential that these efforts be complemented with refined analysis methods that allow mitigation of the such systematics in the observed maps. At the very least, these methods will serve as important null tests, which will need to be performed to claim a robust primordial B-mode signal. 
\par 
This paper is organized as follows:
We begin with a review of different map level instrument systematics in Section~\ref{sec:distorions}, indicating the levels of contamination that may be induced by different types of distortion fields.
We then discuss the details of the respective QE in Section~\ref{sec:qe}. 
The iterative cleaning process which we employ in this work is presented in Section~\ref{sec:iter_clean}. Here we also introduce a semi-analytical forecasting procedure that allows us to predict the expected cleaning of B-mode maps. 
In Section~\ref{sec:T to P Sims} we present details of the simulation where the differential gain systematic is injected and discuss why realistic scan strategy is needed to give credible results.
The results from QE analysis on simulated data are presented in Section~\ref{sec:reconstruction}. We do this for the ideal case to show the limits of the QE method and for the more realistic, non-ideal case.
%
We further quantify the results in Section \ref{sec:r} where we discuss the impact on cleaning on the inferred tensor to scalar ratio $r$ using a likelihood based approach. 

\section{Distortions of the CMB\label{sec:distorions}}
\label{sec:recap}
In an ideal setting the true CMB polarisation signal would be isolated and easily measured without introducing any distortions. However, in practice the measurement are subject to a number of measurement artefacts which need to be controlled and corrected post measurement via some modelling. These contaminants can be typically characterised by their spin dependence and, as such, readily written in to a set of distortion fields.

Both \citep{2010PhRvD..81f3512Y} and \citep{2003PhRvD..67d3004H} use a M\"uller matrix approach \citep{Odea07} to describe the various systematic and cosmological signals that could affect 
measurements of the CMB polarization using a series of distortion fields. These distortions can be written as a coupling between different spin combinations of the instrument and observable fields,
\begin{eqnarray}
    \dX =&&\, [a \pm i 2\omega](\hat{\mathbf{n}}){}_{\pm}\tilde{X}(\uvec{n}) + {}_{\pm}f(\uvec{n}){}_{\mp}\tilde{X}(\uvec{n}) + {}_{\pm}\gamma(\uvec{n})\tilde{T}(\hat{\mathbf{n}})
\,     \label{eqn:Distortion Fields}+\,\sigma \mathbf{{}_{+1}p}(\hat{\mathbf{n}})\cdot {}_{\mp}\eth {}_{\pm}\tilde{X}(\uvec{n};\sigma_{\rm fwhm}) \nonumber\\
    && + \sigma {}_{\pm}d(\uvec{n}) {}_{\pm}\eth\tilde{T}(\hat{\mathbf{n}};\sigma_{\rm fwhm}) +\sigma^{2}q(\hat{\mathbf{n}}){}_{\pm}\eth^2\tilde{T}(\hat{\mathbf{n}};\sigma_{\rm fwhm}) + ...
\end{eqnarray}
where ${}_{\pm}\tilde{X}(\hat{\mathbf{n}}) = \tilde{Q} \pm i\tilde{U}$ is the spin $\pm2$, uncontaminated cosmological polarisation signal, $\tilde{T}$ represents the cosmological temperature signal, and ${}_{\pm}\delta X(\hat{\mathbf{n}})$ denotes the total induced distortion, where ${}_{\pm}\eth$ denote the spin raising/lowering operators respectively. The $a(\hat{\mathbf{n}})$ and $\omega(\hat{\mathbf{n}})$ terms are scalar fields describing an amplitude scaling and a polarisation plane rotation respectively, and ${}_{\pm}f(\uvec{n})$ is a spin $\pm 4$ field which couples the conjugate polarisation fields. The ${}_{\pm}\gamma(\uvec{n})$ field is spin $\pm2$ field and couples the temperature to polarisation directly. $\mathbf{{}_{\pm 1}p}$ is a spin $\pm1$ deflection field that describes direction changes of the photons, ${}_{\pm}d(\uvec{n})$ is a spin $\pm1$ field and $q(\hat{\mathbf{n}})$ is a scalar field that couple the first and second derivatives of the temperature field to the polarisation respectively. Since $\X$ is a spin $\pm2$ field the distortions to this must also be spin $\pm2$. 

Equation \ref{eqn:Distortion Fields} has been constructed such that the top line corresponds to mixing between polarisation and temperature in a known direction on the sky. The bottom line presents terms which involve mixing in a local region of the sky with some directional dependence such that they leak the derivative of the CMB fields, such as a pointing error or lensing. The length scale $\sigma_{\rm fwhm}$ corresponds to the width of a Gaussian beam that is smoothing the CMB fields. The terms in the second line are sourced by a simple first order Taylor expansion of the CMB fields around $\hat{\mathbf{n}}$. 

\begin{figure}[ht!]
    \centering
    \includegraphics[width=1\columnwidth]{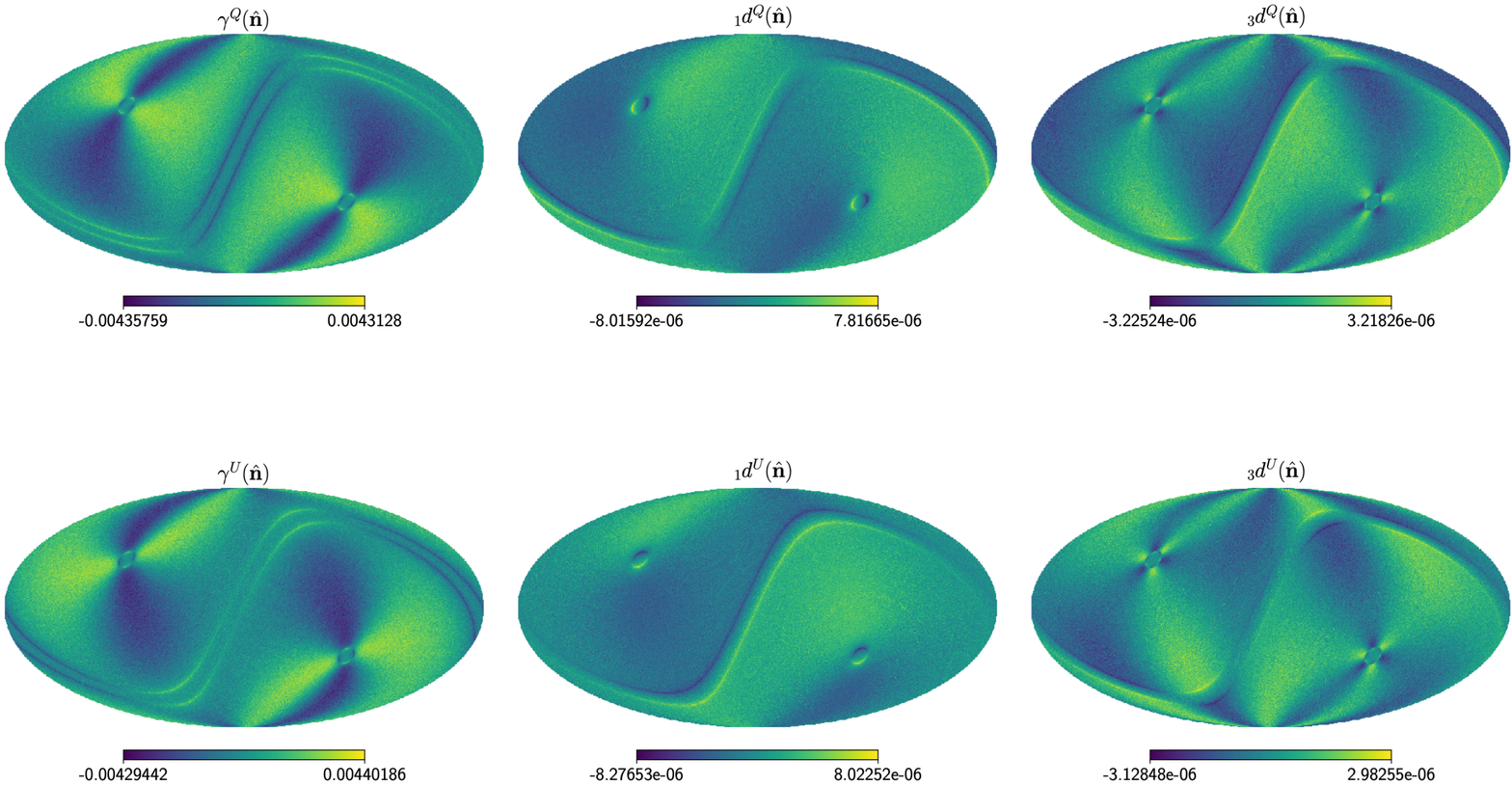}
\caption{The left column shows maps of the spin-2 $(\gamma^{Q}+i\gamma^{U})(\hat{\mathbf{n}})$ distortion field sourced by a 1\% differential gain systematic which causes a T leakage into the polarisation. The second and third columns show the spin-1 $({}_{1}d_{1}+i{}_{1}d_{2})(\hat{\mathbf{n}})$ and spin-3 $({}_{3}d_{1}+i{}_{3}d_{2})(\hat{\mathbf{n}})$ distortion fields sourced by a $0.1'$ differential pointing systematic which causes a leakage of the derivative of the temperature into the polarisation. These levels of gain and pointing are typical of those found in recent CMB experiments \citep[e.g.][]{2015ApJ...814..110B,1403.2369}.} 
   \label{fig:Distortion Fields}    
\end{figure}

There are no known processes that cause T to P conversion along the line of sight. Therefore, each of the $\tilde{T}(\uvec{n})$ containing terms can be attributed to some systematic. 
Some example distortion fields are depicted in Figure~\ref{fig:Distortion Fields}. Here it is important to note that the form of these distortion fields depends on the scanning strategy and therefore a realistic scan should be employed when assessing the importance of different systematics. We will revisit this detail in Section~\ref{sec:Realistic Scan}. 

Note that we only consider CMB fields and their distortions in this work. In particular, we do not include foreground fields; We expect that foregrounds would be removed using standard techniques \cite{Delabrouille2009, 10.1093/ptep/ptu065}, and we leave for future work any complications arising from interactions between the two methods.


%
In this paper we will focus on the T to P leakage mediated by the field ${}_{\pm}\gamma(\uvec{n})$, which is sourced by differential gain variations in the detector \citep{2015ApJ...814..110B,1403.2369} coupled with the instrument scan. Earlier work noted that the ${}_{\pm}\gamma(\uvec{n})$ field poses the largest potential obstruction to the robust primordial B-mode recovery \citep{2010PhRvD..81f3512Y}. Specific details on inclusion of these systematics in the simulated  CMB maps will be presented in Section \ref{sec:T to P Sims}.

\subsection{Additional distortion terms}
\label{sec:Additional Terms}
There exist other systematics that contribute at leading order to the distortions described in Equation~\eqref{eqn:Distortion Fields}. These have been ignored in previous literature \cite{2010PhRvD..81f3512Y}. For completeness, here we briefly discuss these `new' terms; these are encoded in the following expression,
%
\begin{eqnarray}
\label{eqn:Additional Distortion Terms}
    {}_{\pm}\delta X^{\text{New}}(\hat{\mathbf{n}}) &&= \sigma \mathbf{{}_{-1}p}(\hat{\mathbf{n}}) \cdot {}_{\pm}\eth {}_{\pm}\tilde{X}(\hat{\mathbf{n}};\sigma) + \sigma \mathbf{{}_{3}p}(\hat{\mathbf{n}}) \cdot {}_{\pm}\eth {}_{\mp}\tilde{X}(\hat{\mathbf{n}};\sigma) + \sigma \mathbf{{}_{5}p}(\hat{\mathbf{n}}) \cdot {}_{\mp}\eth {}_{\mp}\tilde{X}(\hat{\mathbf{n}};\sigma)
    \nonumber\\
    &&+ \sigma \mathbf{{}_{3}d}(\hat{\mathbf{n}}) {}_{\mp}\eth\,\tilde{T}(\hat{\mathbf{n}};\sigma) + \sigma^{2} \mathbf{{}_{4}q}(\hat{\mathbf{n}}){}_{\mp}\eth^{2}\tilde{T}(\hat{\mathbf{n}};\sigma).
\end{eqnarray}
where all the symbols have the same meaning as before.

In particular, note that a differential pointing systematic, which would contribute to the terms containing the ${}_{\pm}d(\hat{\mathbf{n}})$ and $\mathbf{{}_{1}p}(\hat{\mathbf{n}})$ field in Equation~\eqref{eqn:Distortion Fields}, also contributes to a number of terms in Equation~\eqref{eqn:Additional Distortion Terms}. This includes additional coupling of the conjugate of the derivative of the temperature field with the polarisation through the spin-3 $\mathbf{{}_{3}d}(\hat{\mathbf{n}})$ field. From Figure~\ref{fig:Distortion Fields}, it is clear that this contributes a systematic at a level comparable to the spin-1 contribution, $\mathbf{{}_{1}p(\hat{n})}$, and as such may not be neglected. In addition the differential pointing will also contribute to the terms containing the fields $\mathbf{{}_{-1}p}(\hat{\mathbf{n}})$, $\mathbf{{}_{5}p(\hat{\mathbf{n}})}$, and $\mathbf{{}_{3}p(\hat{\mathbf{n}})}$ which induce polarisation mixing as detailed in \cite{McCallum2020}. 

The bias sourced by these additional systematics could in principle hinder a robust claim of primordial $B$-mode detection. Neglecting these terms can potentially lead to distortion fields sourced by some systematics remaining undiagnosed, but we leave more detailed explorations of these new systematics to future work.

\section{Quadratic estimators\label{sec:qe}}

For a statistically isotropic (SI) CMB sky, the off-diagonal correlation of the harmonic space covariance matrix $\langle X_{lm} X'_{l' m '} \rangle \propto \delta_{ll'} \delta_{mm'}$, for $X,X' \in [T,E,B]$. However, secondary anisotropies and measurement artifacts such as the distortion fields described in Section~\ref{sec:distorions} can induce off-diagonal correlations. Therefore, by measuring and combining optimally combining these off-diagonal correlations it is possible to draw inferences on the fields that induce deviations from SI. This is commonly referred to as the quadratic estimator (QE) technique and it has been very successfully used to measure the subtle signatures of weak lensing in the CMB and deduce the lensing deflection angle map \citep{Carron2017,Sherwin2015,Lewis2006}, to test deviations from the standard cosmological assumption of isotropy \citep{planck2016_si} and also to seek signatures of non-standard physics \citep{Namikawa_2020,PhysRevD.96.102003,Williams_2020}. Some aspects of QE can be discussed quite generally without specific details about the systematics that source the distortions fields, and we refer the reader to \citep{2010PhRvD..81f3512Y} for such a discussion. In this work we focus out attention to the QE required to reconstruct the spin-2 $\gamma$ fields which mediates the T to P leakage, the details of which we discuss next.

\subsection{Quadratic estimator for the spin-2 $\gamma$ field}
We improve on the work presented in \citep{2010PhRvD..81f3512Y}, by first deriving the full sky QE (i.e without making the flat sky approximation) and then presenting its efficient real space form.  We begin by writing the map level model for the observed, distorted CMB sky which is given by the following expression,
\begin{equation}\label{eqn:data model}
    \X =   {}_{\pm}\tilde{X}(\uvec{n})\star B(\uvec{n}) + {}_{\pm}\gamma(\uvec{n})\cdot [\tilde{T}(\uvec{n})\star B(\uvec{n})] + N(\uvec{n})\,,
\end{equation}
 where, $\star$ represents a convolution operation  and $\cdot$ represents a scalar multiplication, $B$ denotes the beam, ${}_{\pm}\gamma(\uvec{n})$ represents the T to P leakage fields, and finally $N(\uvec{n})$ represents the measurement noise.  Given that ${}_{\pm}X$ and ${}_{\pm}\gamma$ are spin two fields, they can be decomposed in the spin weighted spherical harmonic basis as follows,
\begin{subequations}
\begin{eqnarray}
\X&&\,\equiv \sum_{lm}\Xlm{l_{1}m_{1}}\,\Yspin{l_{1}m_{1}}= \Q \pm i\U\;,\\
{}_{\pm}\gamma(\uvec{n})&&\,\equiv \sum_{LM}\GLM\,\Yspin{LM}= \gQ\pm i\gU\;,
\end{eqnarray}
\end{subequations}
where the spherical harmonic coefficients can be expressed in terms of the scalar $E$ and pseudo scalar $B$ as follows,
\begin{subequations} \label{eq:eb_rep}
\begin{eqnarray} 
E_{l_{1}m_{1}}\, = -\frac{1}{2}\left({}_{+}X_{l_{1}m_{1}} + {}_{-}X_{l_{1}m_{1}}\right) &\;;& B_{l_{1}m_{1}}\, = -\frac{1}{2i}\left({}_{+}X_{l_{1}m_{1}} - {}_{-}X_{l_{1}m_{1}}\right) \,, \\
\gELM\, = -\frac{1}{2}\left({}_{+}\gamma_{LM} + {}_{-}\gamma_{LM}\right) &\;;& \gBLM\, = -\frac{1}{2i}\left({}_{+}\gamma_{LM}- {}_{-}\gamma_{LM}\right)\;.
\end{eqnarray}
\end{subequations}
Given these definition the harmonic space coefficients of expansion of the contaminant spin-2 field is given by,
\begin{eqnarray}
    \dist{l_{1}m_{1}}&&\, = \sum_{LM}\sum_{l_{2}m_{2}}\GLM \,\tilde{T}_{l_{2}m_{2}}\int d\uvec{n}\,\Yspin{LM}Y_{l_{2}m_{2}}(\uvec{n}){}_{\pm2}Y^*_{l_{1}m_{1}}(\uvec{n})\;,\nonumber\\
    &&\, =  \sum_{LM}\sum_{l_{2}m_{2}}\GLM \,\tilde{T}_{l_{2}m_{2}}\,\Ipm{Ll_{2}l_{1}}{Mm_{2}m_{1}}\;,\label{eqn:deltaX}
\end{eqnarray}
where that both $\dist{l_{1}m_{1}}$ and $\tilde{T}_{l_{2}m_{2}}$ fields are beam convolved. 
We note that the integral $\Ipm{Ll_{2}l_{1}}{Mm_{2}m_{1}}$ has the property:  $\Iplus{Ll_{2}l_{1}}{Mm_{2}m_{1}} = (-1)^{\ell}\Iminus{Ll_{2}l_{1}}{Mm_{2}m_{1}}$
where $\ell\equiv L+l_{1}+l_{2}$.
Motivated by this property we define the even and odd parity projection operators as: $ P_{e/o}=\frac{\left( 1 \pm (-1)^{\ell} \right)}{2}$, which as we will see allows to condense a lot of the algebra that follows. Given all the definitions, Equation \eqref{eqn:deltaX} can be re-expressed in the following form,
\begin{subequations}
\begin{eqnarray}
    \delta E_{l_{1}m_{1}} &&\,=  -\frac{1}{2}\sum_{LM}\sum_{l_{2}m_{2}}\left({}_{+}\gamma_{LM}\tilde{T}_{l_{2}m_{2}}\,\Iplus{Ll_{2}l_{1}}{Mm_{2}m_{1}} + {}_{-}\gamma_{LM}\tilde{T}_{l_{2}m_{2}}\,\Iminus{Ll_{2}l_{1}}{Mm_{2}m_{1}}\right)\;,\label{eqn:deltaE}\\
    \delta B_{l_{1}m_{1}} &&\,=  \frac{i}{2}\sum_{LM}\sum_{l_{2}m_{2}}\left({}_{+}\gamma_{LM}\tilde{T}_{l_{2}m_{2}}\,\Iplus{Ll_{2}l_{1}}{Mm_{2}m_{1}} - {}_{-}\gamma_{LM}\tilde{T}_{l_{2}m_{2}}\,\Iminus{Ll_{2}l_{1}}{Mm_{2}m_{1}}\right)\label{eqn:deltaB}\;.
\end{eqnarray}
\end{subequations}
The above equations can be further reduced to be expressed in terms of the $\gELM$ and $\gBLM$, which after some simple algebra can be expressed in the following form,
\begin{subequations}
\begin{eqnarray}
    \delta E_{l_{1}m_{1}}&&\, = \sum_{LM}\sum_{l_{2}m_{2}}\left[\gELM \tilde{T}_{l_{2}m_{2}}\,\Iplus{Ll_{2}l_{1}}{Mm_{2}m_{1}} P_e + i\gBLM \tilde{T}_{l_{2}m_{2}} \,\Iplus{Ll_{2}l_{1}}{Mm_{2}m_{1}}P_o\right]\;,\label{eqn:dE without I}\\
    \delta B_{l_{1}m_{1}}&&\, = \sum_{LM}\sum_{l_{2}m_{2}}\left[\gBLM \tilde{T}_{l_{2}m_{2}}\,\Iplus{Ll_{2}l_{1}}{Mm_{2}m_{1}} P_e - i\gELM \tilde{T}_{l_{2}m_{2}} \,\Iplus{Ll_{2}l_{1}}{Mm_{2}m_{1}}P_o\right]\;.\label{eqn:dB without I}
\end{eqnarray}
\end{subequations}
The polarization contamination  in the measured $E$ or $B$ fields generated by the $\gBLM$ and $\gELM$ fields can be treated separately by choosing a specific parity. For example, $ \delta B_{l_{1}m_{1}}$ for the $\ell=\text{even}$ parity is given by,
\begin{equation}\label{eqn:contaminant B}
    \delta B_{l_{1}m_{1}} = \sum_{LM}\sum_{l_{2}m_{2}}\gBLM \tilde{T}_{l_{2}m_{2}}\,\Iplus{Ll_{2}l_{1}}{Mm_{2}m_{1}} P_e\;,
\end{equation}
from which we will derive an estimator that will allow us to reconstruct $\gBLM$. Choosing the odd parity mode for $\delta B$ will allow us to derive a QE that will allow us to reconstruct $\gELM$. Note that QE derivations for other mode combinations follow a near identical procedure. To illustrate the key points we now carry forward the derivation of the QE for $\gBLM$ starting from Equation~\eqref{eqn:contaminant B}.

The cross correlation between the observed temperature anisotropy map with the observed B-mode maps, under an ensemble average is given by the following expression,
\begin{equation}
    \left\langle \bobs{l_{1}m_{1}}(\tobs{l_{1}'m_{1}'})^{*}\right\rangle =  \sum_{LM}\gBLM \tilde{C}^{TT}_{l_{1}'}\,\Iplus{Ll_{1}'l_{1}}{Mm_{1}'m_{1}} P_e\;.\label{eqn:full correlation}
\end{equation}
where we have implicitly assumed that the correlation between the true temperature and the true B-mode map is zero owing to parity arguments\footnote{For this particular TB QE it is important to note that this estimator does not suffer from any mean field bias and therefore we will not address this detail further.}. Throughout this work $\tilde{C}^{TT}_{l}$ represents the beam convolved power spectrum of the primordial CMB temperature signal.
We now introduce another identity (see Appendix~\ref{app:QE appendix} for details) associated with the integral $I$,
\begin{equation}
\sum_{m_{1}m_{2}}\Ipm{Ll_{2}l_{1}}{Mm_{2}m_{1}}\,\Ipm{L'l_{2}l_{1}}{M'm_{2}m_{1}} = \frac{(H^{L}_{l_{2}l_{1}})^{2}}{2L+1}\delta_{LL'}\delta_{MM'}\label{eqn:I identity}\;,
\end{equation}
where $H^{L}_{l_{2}l}$ is defined in terms of Wigner-3j symbol as,
$
    H^{L}_{l_{2}l_{1}}\equiv  \sqrt{\frac{(2L+1)(2l_{2}+1)(2l_{1}+1)}{4\pi}}\left(
    \begin{array}{ccc}
         L & l_{2} & l_{1}  \\
         -2 & 0 & 2\\
    \end{array}
    \right)\;.
$
Using this identity, the estimator for the correlation in equation~\eqref{eqn:full correlation} can be shown to reduce to the following form,
\begin{equation}\label{eqn:full intermediate}
    \sum_{m_{1}m_{1}'}\bobs{l_{1}m_{1}}(\tobs{l_{1}'m_{1}'})^{*}\Iplus{Ll_{1}'l_{1}}{M'm_{1}'m_{1}} = \gBLMest \tilde{C}^{TT}_{l_{1}'}P_{e}\,\frac{(H^{L}_{l_{1}'l_{1}})^{2}}{2L+1}\;.
\end{equation}
Note that the ensemble average from \eqref{eqn:full correlation} is no longer included here. In reality, we only have access to a single realization of the observed polarization fields when estimating $\gBLM$ and $\gELM$. This is also why it is necessary to replace the $\gBLM$ symbol in \eqref{eqn:full correlation} with the symbol for the estimate $\gBLMest$ in \eqref{eqn:full intermediate}.
We can easily invert equation~\eqref{eqn:full intermediate} to construct an estimator for $\gBLM$ given by,
\begin{equation}
    (\gBLMest)_{l_{1}l_{1}'} = \frac{\sum_{m_{1}m_{1}'}\bobs{l_{1}m_{1}}(\tobs{l_{1}'m_{1}'})^{*}\,\Iplus{Ll_{1}'l_{1}}{M'm_{1}'m_{1}}}{F^{L}_{l_{1}'l_{1}}}\;,\label{eqn:gBLM est raw}
\end{equation}
where $F^{L}_{l_{1}'l_{1}}\equiv \tilde{C}^{TT}_{l_{1}'}P_{e}\,\frac{(H^{L}_{l_{1}'l_{1}})^{2}}{2L+1}$.
This however is only an estimator for  $\gBLM$ from a single multipole pair $(l,l')$. It is now possible devise a minimum variance estimator (MVE) by carrying out the inverse variance weighted sum of the estimator across all possible multipole pairs. For this purpose we begin by evaluating the variance of the estimator for a given multipole pair and this is given by the following expression,
\begin{eqnarray}
\mathcal{C}^{L}_{l_{1}l_{1}'}\equiv\left\langle\gBLMest(\widehat{\gamma}^{B}_{L_{2}M_{2}})^{*}\right\rangle &&\,= \sum_{m_{1}m_{1}'}\sum_{m_{2}m_{2}'}\frac{\bobs{l_{1}m_{1}}(\tobs{l_{1}'m_{1}'})^{*}(\bobs{l_{2}m_{2}})^{*}\tobs{l'_{2}m'_{2}}\,\Iplus{Ll_{1}'l_{1}}{M'm_{1}'m_{1}}\,\Iplus{L_{2}l'_{2}l_{2}}{M'_{2}m'_{2}m_{2}}}{F^{L}_{l_{1}'l_{1}}F^{L_{2}}_{l_{2}'l_{2}}}\;,\nonumber\\
&&\,=\frac{(2L+1)}{(H^{L}_{l_{1}'l_{1}})^{2}}\frac{\widehat{C}^{BB}_{l_{1}}\widehat{C}^{TT}_{l_{1}'}}{\tilde{C}^{TT}_{l_{1}'}\tilde{C}^{TT}_{l_{1}'}(P_{e})^{2}}\;.\label{eqn:gammaB var}
\end{eqnarray}
The $\widehat{C}^{TT}_{l_{1}}$ and $\widehat{C}^{BB}_{l_{1}}$ terms are the power spectra estimated from the observed temperature and $B$-mode polarization fields respectively

Performing an inverse variance weighted sum of the estimator in \eqref{eqn:gBLM est raw} over $l_{1}l_{1}'$ yields the  MVE QE,
\begin{equation}
\gBLMest = N^{\gamma^{B}}_{L}\sum_{l_{1}m_{1}}\sum_{l_{1}'m_{1}'}\frac{\bobs{l_{1}m_{1}}(\tobs{l_{1}'m_{1}'})^{*}\tilde{C}^{TT}_{l_{1}'}\Iplus{Ll_{1}'l_{1}}{Mm_{1}'m_{1}}P_{e}}{\widehat{C}^{BB}_{l_{1}}\widehat{C}^{TT}_{l_{1}'}}\;,\label{eqn:gamma B}
\end{equation}
where $N^{\gamma^{B}}_{L}$ is a normalization, which is also the reconstruction noise (i.e. the power spectrum of the noise in the reconstructed $\gBLM$ map) which is given by inverse of the variances of all the modes added in parallel, specifically,
\begin{equation}
    N^{\gamma^{B}}_{L} = \left[\sum_{l_{1}l_{1}'}\frac{(H^{L}_{l_{1}'l_{1}})^{2}}{(2L+1)}\frac{(\tilde{C}^{TT}_{l_{1}'}P_{e})^{2}}{\widehat{C}^{BB}_{l_{1}}\widehat{C}^{TT}_{l_{1}'}}\right]^{-1}\;.\label{eqn:gammaB recon}
\end{equation}
Note that in Equation~\eqref{eqn:gamma B} and Equation~\eqref{eqn:gammaB recon} only the even parity modes (i.e. $L+l_1 + l_1' \rightarrow \text{Even}$) contribute, which only corresponds to only half the elements in the harmonic space covariance matrix. One can show that the other half of the $TB$ harmonic space covariance matrix encodes information on the E-modes of the spin-2 $\gamma$ field. Following the same procedure as described above, considering the $\ell=\text{Odd}$ modes, it can be shown that the estimator for $\gELM$ is given by the following expression,
\begin{equation}
    \gELMest = -iN^{\gamma^{E}}_{L}\sum_{l_{1}m_{1}}\sum_{l_{1}'m_{1}'}\frac{\bobs{l_{1}m_{1}}(\tobs{l_{1}'m_{1}'})^{*}\tilde{C}^{TT}_{l_{1}'}\Iplus{Ll_{1}'l_{1}}{Mm_{1}'m_{1}}P_{o}}{\widehat{C}^{BB}_{l_{1}}\widehat{C}^{TT}_{l_{1}'}}\;,\label{eqn: gamma E}
\end{equation}
with the reconstruction noise, analogously given by the following expression,
\begin{equation}
      N^{\gamma^{E}}_{L} = \left[\sum_{l_{1}l_{1}'}\frac{(H^{L}_{l_{1}'l_{1}})^{2}}{(2L+1)}\frac{(\tilde{C}^{TT}_{l_{1}'}P_{o})^{2}}{\widehat{C}^{BB}_{l_{1}}\widehat{C}^{TT}_{l_{1}'}}\right]^{-1}\;.\label{eqn:gammaE recon}
\end{equation}
While the forms of the reconstruction noise for $\gBLMest$ and $\gELMest$ are nearly the same, they differ in the parity of modes that contribute to the sum and therefore their numerical values are not identical values. These are curved sky equivalents of the flat sky estimators presented in Equation (10) of \cite{2010PhRvD..81f3512Y}. 
\subsection{The reconstruction noise\label{sec:recon noise}}
When reconstructing the distortion fields, in our case $\gELM$ and $\gBLM$, the reconstruction noise determines which harmonic modes of these fields can be recovered. Multipoles that are dominated by reconstruction noise cannot be properly reconstructed. Therefore, it is important to perform the reconstruction with as little noise as possible. While the reconstruction noise can be generally reduced by decreasing the measurement noise and increasing the angular resolution of the measurements, we will be interested in minimizing the reconstruction noise for a fixed instrument configuration.
\begin{figure}[t!]
    \centering
    \begin{subfigure}[t]{0.49\textwidth}
    \centering
    \includegraphics[width=1\linewidth]{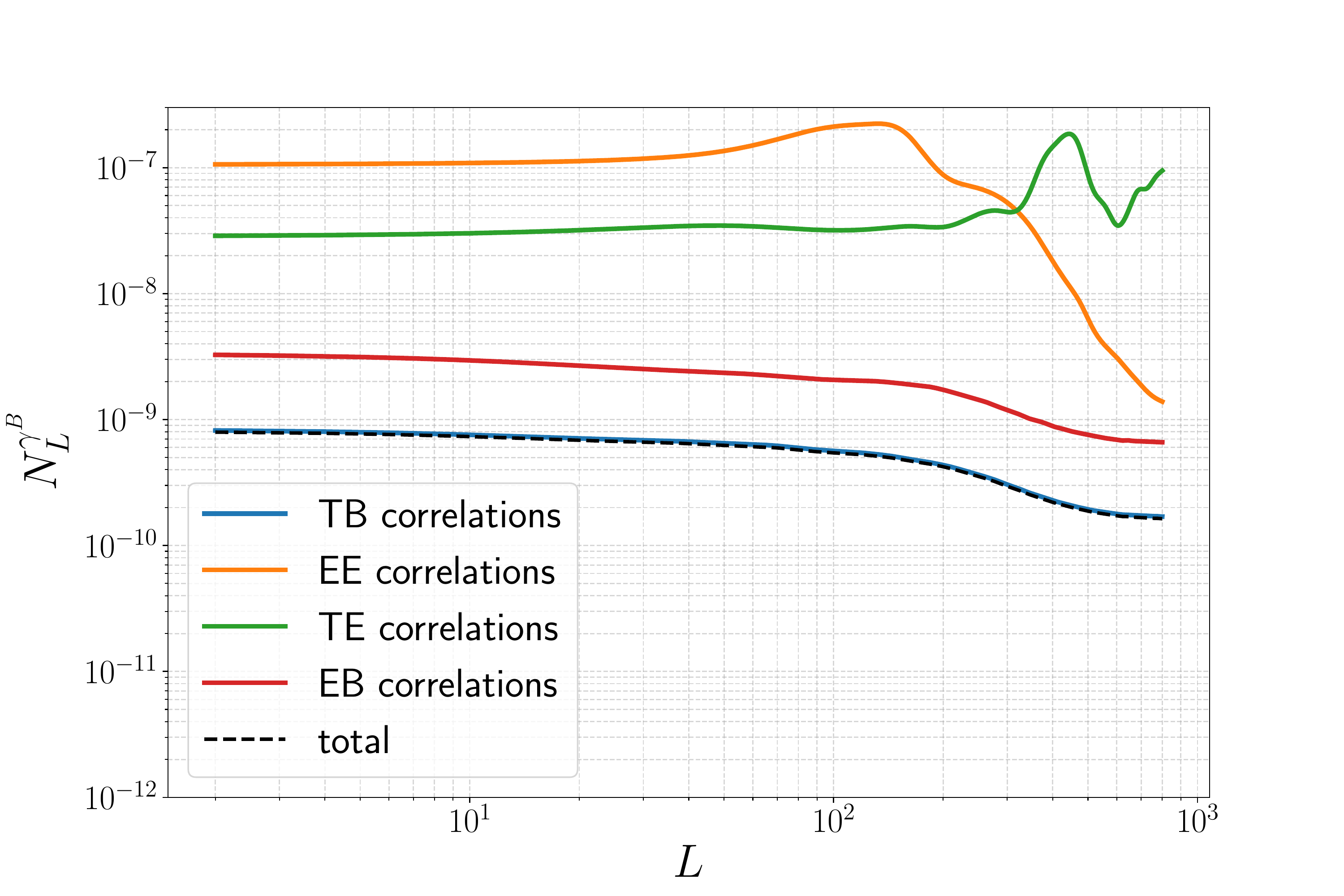}
    \end{subfigure}
    \hfill
    \begin{subfigure}[t]{0.49\textwidth}
    \centering
    \includegraphics[width=1\linewidth]{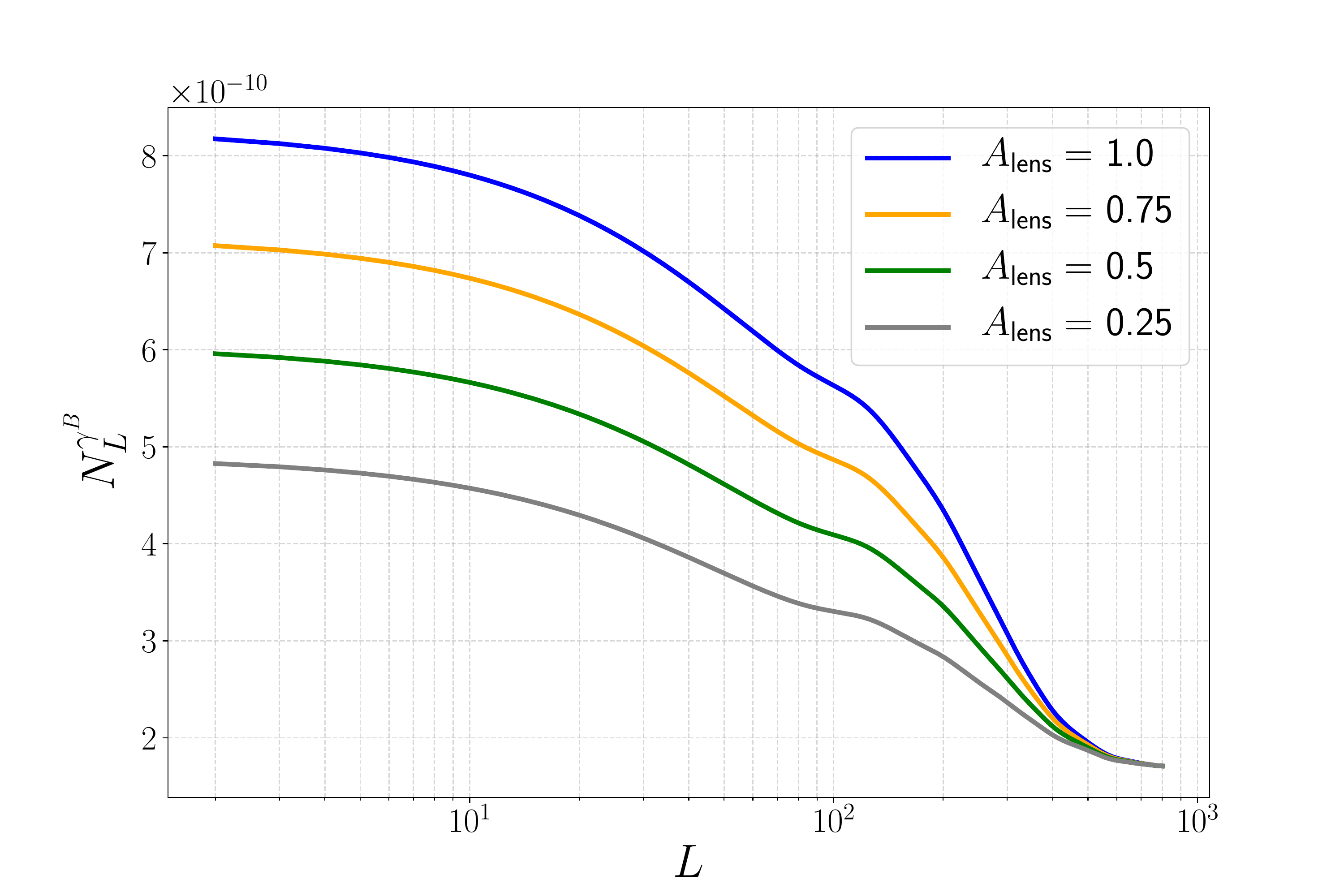}
    \end{subfigure}   
   
    \caption{The left panel shows the reconstruction noise
estimated for the respective QE assuming $A_{\text{lens}} = 1$. The reconstruction noise for the combination of all the estimators is indicated with a black dashed line. The right panel shows the how the TB reconstruction
noise varies as a function of $A_{\text{lens}}$. Both plots assume instrument noise and
beam characteristics compatible with the \emph{LiteBIRD} instrument. Here we display only $N^{\gamma^{B}}_{L}$, as $N^{\gamma^{E}}_{L}$
shows the same trends.  \label{fig:mode choice}}
\end{figure}
The reconstructions can be performed using QE constructed from a variety of cross correlations: $EE$, $TE$, $TB$ \& $EB$ and by optimally combining all of them.  The reconstruction noise resulting from combining all the estimators naturally is the best. However this requires significant more effort which may not be warranted.  This can be assessed by estimating the reconstruction noise for each of the estimators and performing a relative comparison. 

We perform this test in three different settings: (i) in the cosmic variance limited ideal case, (ii) in the more realistic, non-ideal case with a noise sensitivity of $w^{-1}_{TT}=2.7\mu \text{K}\,\text{arcmin}$ and a full-width at half-maximum of $\theta_{\text{FWHM}}=30^{\prime}$,  and (iii) assuming the CMBpol configuration with $w^{-1}_{TT}=1.4\mu \text{K}\,\text{arcmin}$ and $\theta_{\text{FWHM}}=4^{\prime}$. The results from this exercise, for the $w^{-1}_{TT}=2.7\mu \text{K}\,\text{arcmin}$, and $\theta_{\text{FWHM}}=30^{\prime}$ case study, are summarized in Figure~\ref{fig:mode choice}. In all three cases we find the reconstruction noise associated with the TB QE is the lowest and closely matches the estimated reconstruction noise expected from optimally combining all the different QE. Motivated by this observation we derive all our results using only the TB estimator.

These findings differ from the conclusions drawn in \citep{2010PhRvD..81f3512Y} where it is stated that for the CMBpol configurations the EB QE yields the lowest reconstruction noise. This highlights the importance of using detailed case studies for specific distortion fields in order to test the conventional wisdom when applying quadratic estimators. The highest sensitivity choice of correlations for any given distortion field depends on the details of the estimator and it is therefore important to check which choice of correlations yields the most desirable reconstruction noise for each distortion field separately.

\subsubsection{Impact of de-lensing on reconstruction noise}
The B-modes generated by weak lensing of the CMB act as a competing signal for measurement of primordial B-mode signal sourced by tensor perturbations. Therefore many upcoming analysis strategies necessarily include a de-lensing procedure, thereby reducing the additional variance introduced by the lensing signal and consequently improving the measurements or upper bounds on $r$. CMB $B$-mode power induced by lensing also contributes to the reconstruction noise for the TB correlations. The blind systematic cleaning being proposed in this work can also potentially\footnote{In principle there might be an additional coupling between the lensing and distortion fields which may bias the lensing reconstruction. We leave the exploration of this subtlety to future work.} benefit from the reduced impact of lensing on the observed $B$-mode spectrum. Note that in this work we do not carry out the de-lensing procedure, but model it simply by scaling the lensing power spectrum with an amplitude $A_{\rm lens}$.

In the cosmic variance limited case the reconstruction noise scales very simply with different amounts of delensing such that $N^{\gamma}_{L}= \left.A_{\text{lens}}N^{\gamma}_{L}\right|_{A_{\text{lens}=1}}$. This simple relationship breaks down in a realistic scenario where the dominant contribution to the $B$-mode variance at high $l$ comes from the instrument noise. This results in de-lensing having little impact on the reconstruction noise for high $L$ modes. \textbf{In the more realistic, non-ideal case} this translates to modes greater than $L\sim 600$, as is seen in Figure~\ref{fig:mode choice}. De-lensing can still reduce the $B$-mode variance for modes where the instrument noise is not the dominant contribution, and this will be encoded as a reduction in reconstruction noise for lower multipoles. 


%
It is possible to further minimise the impact of the reconstruction noise by carrying out iterative cleaning of the CMB maps and by constructing the optimal filter for $\gELMest$ and $\gBLMest$ and we will return to discussing this in Section~\ref{sec:iter_clean}.

\subsection{Efficient real space estimators}
These harmonic space estimators derived above involve large sums over multipoles (scaling roughly as $\sim (l_{\text{max}})^{4}$. and also requires evaluation of the Wigner symbols, therefore they are not very computationally efficient. These estimators appear as convolutions in harmonic space and one therefore expects to be able to express them as direct products of some real space fields. This procedure exists and is routinely implemented for the weak lensing estimators \cite{Hu2000}. Here we derive an analogous real space operator for reconstructing the spin-2 $\gamma$ field. By using the explicit integral form of $\Ipm{Ll^{\prime}_{1}l_{1}}{Mm^{\prime}_{1}m_{1}}$ we can rewrite Equation~\eqref{eqn:gamma B} and Equation~\eqref{eqn: gamma E}, which after some algebra can be expressed in the following form,
\begin{multline}
\gBLMest = \frac{N^{\gamma^{B}}_{L}}{2}\int d\uvec{n}\big[\left({}_{+2}A_{B^{*}}(\uvec{n})\right)^{*}A^{TT}_{T^{*}}(\uvec{n}){}_{+2}Y_{LM}(\uvec{n}) \\+ 
    \left({}_{-2}A_{B^{*}}(\uvec{n})\right)^{*}A^{TT}_{T^{*}}(\uvec{n}){}_{-2}Y_{LM}(\uvec{n})\big]\;,
\end{multline}
\begin{multline}
\gELMest = \frac{-iN^{\gamma^{E}}_{L}}{2}\int d\uvec{n}\big[\left({}_{+2}A_{B^{*}}(\uvec{n})\right)^{*}A^{TT}_{T^{*}}(\uvec{n}){}_{+2}Y_{LM}(\uvec{n}) \\-
    \left({}_{-2}A_{B^{*}}(\uvec{n})\right)^{*}A^{TT}_{T^{*}}(\uvec{n}){}_{-2}Y_{LM}(\uvec{n})\big]\;.
\end{multline}
where $*$ indicates the complex conjugate and the real space fields are defined as follows, 
\begin{eqnarray}
    {}_{\pm2}A_{B^{*}}(\hat{\mathbf{n}}) =&&\, \sum_{l_{1}m_{1}}\frac{1}{\widehat{C}^{BB}_{l_{1}}}(B^{\rm obs}_{l_{1}m_{1}})^{*}{}_{\pm2}Y_{l_{1}m_{1}}(\hat{\mathbf{n}})\;,\\
    A^{TT}_{T^{*}}(\hat{\mathbf{n}}) =&&\, \sum_{l_{1}m_{1}}\frac{\tilde{C}_{l_{1}}^{TT}}{\widehat{C}^{TT}_{l_{1}}}(T^{\rm obs\,'}_{l_{1}m_{1}})^{*}Y_{l_{1}m_{1}}(\hat{\mathbf{n}})\;,
\end{eqnarray}
where $X,Y\in[T,E,B]$. Since these real space fields can be computed independently and merely involve a few spin harmonic transforms, as opposed to explicit multipole sums and evaluations of Wigner symbols, these are significantly more numerically efficient.
%

\section{Iterative cleaning}\label{sec:iter_clean}
The QE technique detailed in the previous section provides an excellent tool for diagnosis and reconstruction of potential contaminants, which we demonstrate in Section~\ref{sec:reconstruction}. However, in this section we shift our attention to discussing how the reconstructed distortion fields can be used to optimally de-contaminate the observed CMB maps. A cleaned CMB map has a lower reconstruction noise, which in effect allows for uncovering the components of the distortions fields that were noise dominated in the original map. These additional components of the reconstructed distortion fields can then be fed back to the cleaning algorithm. This translates to the cleaning of additional contaminated modes. We will refer to this procedure as iterative cleaning; this procedure allows for a more detailed recovery of the distortion fields. We now discuss how the reconstructed distortion fields can be optimally combined with the contaminated maps to yield de-contaminated maps.
\subsection{Optimally de-contaminating the CMB maps\label{sec:cleaning scheme}}
The residual contamination in the de-contaminated B-mode map is given by,
\begin{equation}\label{eqn:B clean full}
    B^{{\rm res}}_{lm} = \delta B_{lm} - \delta\widehat{B}_{lm}\;.
\end{equation}
where $\delta B$ denotes the true contamination and $\delta\widehat{B}$ is the estimated map of contamination.  $\delta\widehat{B}$ sourced by the B-mode component of the $\gamma$ field can be estimated using the following expression,
\begin{equation}\label{eqn:contaminant B f}
    \delta \widehat{B}_{l_{1}m_{1}} = \sum_{LM}\sum_{l_{2}m_{2}}\widehat{\gamma} ^{B}_{LM}\tilde{T}_{l_{2}m_{2}}\Iplus{Ll_{2}l_{1}}{Mm_{2}m_{1}}\eff\;,
\end{equation}
which is the same as Equation~\eqref{eqn:dB without I}, except that the distortion field is replaced by the estimated distortion field using the QE as described in Section~\ref{sec:qe} and we have introduced the weights $\eff$ which need to be determined. 
We define an optimal cleaning algorithm as one that minimizes the power spectrum of the residual contamination after each iteration of cleaning. Given Equation~\eqref{eqn:contaminant B f}, the angular power spectrum of the residual contamination map is given by the following expression,
\begin{equation} \label{eq:res_ps}
    C^{BB,\,\rm res}_{l_{1}} = \sum_{Ll_{2}}\frac{(H^{L}_{l_{2}l})^{2}}{(2l_{2}+1)}\left[\tilde{C}^{\gamma^{B}\gamma^{B}}_{L}\tilde{C}^{TT}_{l_{2}}-2\tilde{C}^{\gamma^{B}\gamma^{B}}_{L}\tilde{C}^{TT}_{l_{2}}\eff + \widehat{C}^{\gamma^{B}\gamma^{B}}_{L}\widehat{C}^{TT}_{l_{2}}(\eff)^{2}\right]\;,
\end{equation}
and the optimal $\delta\widehat{B}$ can be estimated by solving for the weights $\eff$ that minimize Equation~\ref{eq:res_ps}. Taking the derivative with respect to $\eff$ to calculate the minimum of the residual, $C^{BB,\,\rm res}_{l}$, results in a filter of the form,
\begin{equation}
    \eff = \frac{\tilde{C}^{\gamma^{B}\gamma^{B}}_{L}\tilde{C}^{TT}_{l_{2}}}{C^{\widehat{\gamma}^{B}\widehat{\gamma}^{B}}_{L}\widehat{C}^{TT}_{l_{2}}}\;.
\end{equation}
It is useful to think of this filter as being composed of two separable parts $\eff = f^{\gamma^{B}}_{L} f^{T}_{l_{2}}$, where
\begin{equation}
    f^{\gamma^{B}}_{L} =\; \frac{\tilde{C}^{\gamma^{B}\gamma^{B}}_{L}}{\widehat{C}^{\gamma^{B}\gamma^{B}}_{L}}\; ~~ ;~~
    f^{T}_{l_{2}} =\; \frac{\tilde{C}^{TT}_{l_{2}}}{\widehat{C}^{TT}_{l_{2}}}\;,
\end{equation}
which can be understood as being the corresponding Wiener filters for $\gBLM$ and $T_{lm}$ fields, on noting that $\hat{C}_l =\tilde{C}_l  + N_l$.  A similar calculation can be carried through for estimating the contamination sourced by the E-mode component of the $\gamma$ field. This parallels closely the algorithm followed in de-lensing of the CMB sky \cite{Sherwin2015}.

Given the Wiener filtered maps $\widehat{\gamma}^{E, \rm WF}$, $\widehat{\gamma}^{B, \rm WF}$ and $T^{\rm WF}$, the decontaminated polarization maps are given by the following estimator,
 \begin{equation}\label{eqn:cleaning seperate}
    {}_{\pm}X^{{\rm clean},\,i}(\uvec{n}) = {}_{\pm}X^{i}(\uvec{n}) - {}_{\pm}\widehat{\gamma}^{{\rm WF},\,i}(\uvec{n})T^{{\rm WF},\,i}(\uvec{n})\;,
\end{equation}
where we have again used the intergral form of ${}_{\pm}I$ to express the multipole sum in Equation~\ref{eqn:contaminant B f} in its equivalent and efficient real space form.
Note that throughout this derivation we work with the beam convolved fields. 

The index `i' in the above equation indicates the cleaning iteration. For each iteration of the cleaning beyond the zeroth, the cleaned polarization fields from the previous iteration become the new observed fields to be passed to the QE as well as the cleaning estimator. As expected, the temperature field remains unaltered through this cleaning process. Note that since the reconstructed $\gamma$ fields and the corresponding reconstruction noise estimates are continuously updated, the Wiener filters must be freshly estimated at each iteration. This cleaning process is repeated until the reconstruction noise and the B-mode power spectrum converges to their respective floors.
We reiterate that in this case study we focus on the details of the iterative cleaning algorithm for the T to P leakage distortion sourced by differential gain, however, this can be generalized to the full range of distortions described in Section~\ref{sec:distorions}.

\subsubsection{Gaussian Filters}
 While it is important that the filters lead to the smallest residual contamination after each iteration of the cleaning, it is also important that the filter prevents the cleaning process from introducing excess bias. While we have shown that the Wiener filters are the optimal filters that minimize $C^{BB,\,\rm res}_{l}$,  in our numerical experiments working with idealized low noise simulations we find that Wiener filters tend to overestimate the contamination for modes where the reconstruction noise is high, thereby making the iterative procedure have an undesirable non-convergent behaviour. We understand this to be a special feature of a T to P leakage systematic in which $T >> B$ and therefore even a small error in the reconstructed $\gamma$ maps can lead to a large errors in the de-contaminated the B-mode maps in particular. To prevent this we propose a Gaussian filtering scheme,
\begin{equation}
    f^{\gamma}_{L} =  A\exp\left(-\left[\frac{\widehat{C}^{\gamma\gamma}_{L}}{\widehat{C}^{\gamma\gamma}_{L}-N^{\gamma}_{L}}\right]^{2}\right)\;,
\label{eqn:filter}
\end{equation}
where the normalization $A$ is set such that  ${\rm max}(f^{\gamma}_{l})=1$. 
In our numerical simulations we perform, we find this to be a convergent scheme in all cases (unlike Wiener filtering), as it is more aggressive in suppressing modes that are contaminated by noise, thereby preventing excess bias from being introduced into the cleaning.

\subsection{Forecasting the reconstruction noise and $C^{BB}_{l}$ floors\label{sec:iter_forecast}}
Due to the imperfect reconstruction of the $\gamma$ fields, it is in practice not possible to perfectly decontaminate the polarization maps using this procedure. To answer this question we have devised a forecasting procedure that enables us to predict the reconstruction noise and $C^{BB}_{l}$ floors that the iterative cleaning procedure should in principle achieve.
\par 
Making these forecasts involves evaluating the following algorithm. We begin by making an estimate of the reconstruction noise under the assumption that polarization map can be perfectly cleaned. After this initialization we iterate over the following steps until convergence:
\begin{itemize}
\item Use the estimated reconstruction noise to simulate Weiner filtered $\gamma$ maps, using the true systematic maps as input. 
\item Use the filtered $\gamma$ maps to perform cleaning on a simulation of contaminated polarization maps using the same procedure prescribed in Section~\ref{sec:cleaning scheme}.
\item Use the mock cleaned maps to make revised estimates of the reconstruction noise.
\end{itemize}
The $B$-mode calculated from the mock cleaned maps, and the estimates of the reconstruction noise were found to converge after five iterations of the above process. This procedure provides a forecast for both the reconstruction noise floor as well as the cleaned power spectrum characterizing the polarization maps. We can compare these estimates to the reconstruction noise and the polarization power spectra derived from employing the iterative cleaning procedure to assess if the blind cleaning is performing as expected.

Note that this procedure uses information from the true CMB and systematics maps and as such is only useful for testing the analysis pipeline. For an actual experiment, where we can assume no prior knowledge of the contaminants, we will not have the liberty of carrying out such validation tests. For actual data analysis we would carry out iterations until the reconstruction noise converges, as we will demonstrate in Section~\ref{sec:reconstruction}. 

\section{\label{sec:T to P Sims}Simulating Temperature to Polarization Leakage}
\subsection{Systematic - Differential Gain\label{sec:Differential Gain}}

In upcoming experiments, both satellite- and ground-based, control of T to P leakage will be essential. The relative amplitudes of the signals means that even temperature leakage at the percent level could be a significant contaminant to the B-mode signal. This section will describe the simulation of T to P leakage and the scan strategy we consider for our differential gain case study, following the approach of \cite{Wallis2016,McCallum2020}; see those works for a more exhaustive discussion. We reiterate that the quadratic estimator approach is applicable to a wide range of distortions, we are just choosing this systematic for our detailed case study as it generates the $\gamma$ distortion field that was found to be most important in \cite{2010PhRvD..81f3512Y}.
\par
The signal observed by a single detector contains both the temperature and modulated polarisation signals and may be written as
\begin{equation}
    d^X =  (1+\delta g^X)\tilde{T}(\hat{\mathbf{n}}) + \tilde{Q}(\hat{\mathbf{n}})\cos(2\psi) + \tilde{U}(\hat{\mathbf{n}})\sin(2\psi)\;,
    \label{eqn:singledetect}
\end{equation}
where $X$ denotes which detector is being considered, and $\psi$ is the crossing angle (the orientation of the focal plane on the sky). We have represented a possible constant gain or calibration factor by $\delta g^X$, but for our case study we only apply this factor to the temperature signal. A calibration or gain difference can also cause other effects, such as an amplifcation of the polarisation signal, but we focus on this as it is the most significant problem and is the one that can manifest as a $\gamma$ type distortion.

We consider a pair-differencing experiment, consisting of co-located detector pairs that are oriented 90 degrees apart. whose observed signals are differenced. Ideally the temperature signal would be completely removed by this procedure, however, any mismatch in the detector gain $\delta g^X$ between the two detectors will result in leakage of the temperature signal into the polarisation map. We write the differenced signal for a detector pair $i$ as 
\begin{eqnarray}
    d_i =&& \frac{1}{2}\left(d^A_i - d^B_i\right)
    \nonumber\\
    =&&\frac{1}{2}\bigg[(\delta g^A_i-\delta g^B_i)\tilde{T}(\hat{\mathbf{n}}) +2 \tilde{Q}(\hat{\mathbf{n}})\cos(2\psi)+\,2 \tilde{U}(\hat{\mathbf{n}})\sin(2\psi)\bigg]\;,
\end{eqnarray}
where $A$ and $B$ denote each detector in the co-located pair. The temperature leakage will occur if $\delta g_i=\delta g^A_i-\delta g^B_i\neq0$, and is given by 
\begin{equation}
    \delta d^g_i = \frac{\delta g_i}{2} \, \tilde{T}(\hat{\mathbf{n}})\;,
    \label{eqn:DiffGain}
\end{equation}
which is a spin-0 quantity. This will combine with the spin-2 part of the scan strategy to create a spurious spin-2 signal that contaminates the polarisation measurement. We may describe a given scanning strategy by a real space field (again, see \cite{Wallis2016,McCallum2020} for more details)
\begin{equation}
    \tilde{h}_k = \frac{1}{N_{\text{hits}}(\theta, \phi)} \sum_{j} e^{ik\psi_j(\theta, \phi)}\;,
    \label{eqn:h_k}
\end{equation}
where $\psi_j$ is the $j$th crossing angle of a given pixel, and $N_{\text{hits}}$ is the total number of measurements in that pixel. The survey mask is described by $\tilde{h}_0$, while $\tilde{h}_2$ and $\tilde{h}_4$ naturally appear in simple map-making, and various $k$ values contribute to different systematic effects (see e.g. \cite{2009arXiv0906.1188B,Wallis2016,McCallum2020}. In the differential gain case considered here, the systematic couples to $\tilde{h}_2$ \cite{Wallis2016,McCallum2020}, so the spurious signal due to a detector pair $i$ is
\begin{equation}
    {}_2(\delta d^g_i) = \frac{1}{2} \tilde{h}_2(\hat{\mathbf{n}}) (\delta g_i) \, \tilde{T}(\hat{\mathbf{n}})\;.
\end{equation}
In particular for our simulations, we will use a focal plane with two such pairs of detectors, oriented at 45 degrees to one another to allow simultaneous measurement of both $\tilde{Q}$ and $\tilde{U}$ signals. The combined systematic contribution from the two detector pairs is given by
\begin{equation}
{}_2(\delta d^g) = \frac{1}{2} \tilde{h}_2(\hat{\mathbf{n}}) (\delta g_1 - i\delta g_2) \, \tilde{T}(\hat{\mathbf{n}})
\label{eqn:Spin2Signal}
\end{equation}
where the factor of $i$ in the second detector term is due to the rotated orientation of 45 degrees.

For simplicity we choose a setup where each pair of the detectors experiences the same differential gain, $\delta g_1 = \delta g_2 = 10^{-2}$, which corresponds to a 1\% differential gain. This simplification will not affect the generality of the QE results presented, but one may expect slightly different levels of gain mismatch for different focal plane setups. \footnote{There are other methods for handling differential systematics \citep[e.g.][]{bicep2015a,Miller2009}, however here we study the QE approach in detail rather than performing a comparison of methods. One alternative method being investigated is rotating half wave plates, in which case it isn't clear whether detector differencing should be used \cite{Kusaka2014,kusaka2018}. We note that the important systematics in a HWP setup are likely to be different, and it is unclear to what extent our distortion field setup will capture the important systematics. We leave the investigation of the utility of QE in such a situation to future work.}

By comparing equation~\eqref{eqn:Distortion Fields} to \eqref{eqn:Spin2Signal}, we can see that the differential gain be related to the $\gamma$ distortion as
\begin{equation}
\begin{split}
(\gamma^{Q}\pm i\gamma^{U})(\hat{\mathbf{n}})\Tilde{T}(\hat{\mathbf{n}}) = \frac{1}{2} \tilde{h}_{\pm 2}(\hat{\mathbf{n}}) (\delta g_1 \mp i\delta g_2) \, \tilde{T}(\hat{\mathbf{n}})\;.
\end{split}
\label{eqn:Signal}
\end{equation}

\subsection{Differential gain $r$-bias} 
\begin{figure}[t!]
    \centering
    \includegraphics[width=0.9\textwidth]{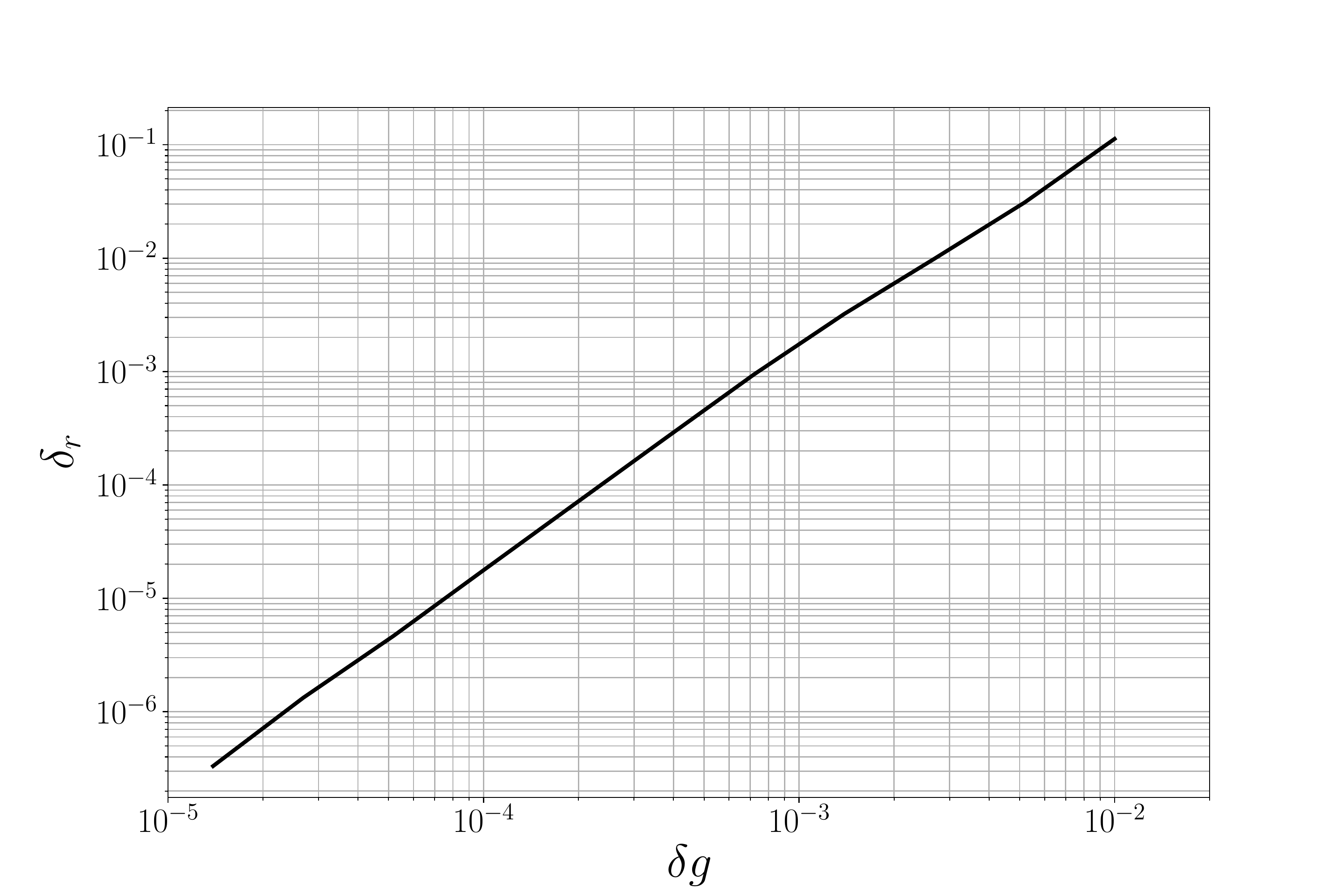}
    \caption{The bias on a fiducial tensor-to-scalar ratio of $r=10^{-3}$ for levels of differential gain of $\delta g = 10^{-6}\rightarrow10^{-2}$ in the presence of a white noise level of $w^{-1}_{TT}=2.7\mu \text{K}\,\text{arcmin}$ and a full-width at half-maximum of $\theta_{\text{FWHM}}=30^{\prime}$. For values of $\delta g \sim 1.4\times 10^{-5}$ the bias is $\delta_r=0$ to machine precision.}
    \label{fig:r bias x gain}
\end{figure}

We examine the expected biases on $r$ that correspond to a range of levels of differential gain. We define the bias sourced by the systematic, $\delta_{r}$, as the difference between the mean of the posterior of the contaminated and uncontaminated spectra. In Figure~\ref{fig:r bias x gain} we show the bias on $r$, $\delta_{r}$, for differential gains in the range $\delta g = 10^{-5} \rightarrow 10^{-2}$, in the presence of a noise sensitivity of $w^{-1}_{TT}=2.7\mu \text{K}\,\text{arcmin}$ and a full-width at half-maximum of $\theta_{\text{FWHM}}=30^{\prime}$. We calculate the biases for a fiducial tensor-to-scalar ratio of $r=10^{-3}$. 
For $\delta g \lesssim 7\times 10^{-4}$ the bias is smaller than the $1\sigma$ statistical variance on $r$ and cleaning would not be necessary. For values greater than this the bias becomes significant, increasing to $\sim 100$ times the fiducial $r$ value for $\delta g \sim 10^{-2}$. In this range the bias will have a significant impact on the robustness of attempts to measure $r$. While the QE cleaning process we present here is able to remove the bias for a range of levels of differential gain, we present results for $\delta g = 10^{-3}$ in order to demonstrate that it is possible to remove even very large levels of contamination sourced by differential detector gain. Note that this larger level of differential detector gain is typical for a number of contemporary ground-based CMB experiments \citep[e.g.][]{2015ApJ...814..110B,1403.2369}, for which we also expect the QE approach to be valuable.

\subsection{Scan Strategy}
\label{sec:Scan Strategy}
We choose to adopt the Experimental Probe of Inflationary Cosmology (\textit{EPIC}) satellite scan strategy \citep{2009arXiv0906.1188B}. This will be representative of other future CMB satellite surveys. The design of the EPIC scan strategy optimises crossing angle coverage and is defined by its boresight angle ($50^\circ$), precession angle ($45^\circ$), spin period (1 min), and precession period (3 hrs) (for further details see \cite{Wallis2016} and \cite{2008arXiv0805.4207B}). This scan is represented as a list of hits, i.e. datapoints, where each hit is specified by its location on the sky (RA and Dec) and parallactic angle ($\psi$). We expect the QE technique to be equally useful for ground based CMB surveys however, because of the role of the scan strategy in the simulations, we have used a satellite survey in this work for two reasons. Firstly, because no ground-based survey covers the whole sky, and we wanted to avoid complications due to partial sky coverage. Secondly, because ground based surveys differ more between experiments and cannot be simply described by a few parameters as satellite scans can, since they depend on complicated constraints and detailed scheduling choices.\footnote{Although see \cite{thomas2021sky} for some simple approaches to approximate ground-based scan strategies that capture the features relevant for studies such as this one.}

\begin{figure}[ht!]
    \centering
    \includegraphics[width=1\columnwidth]{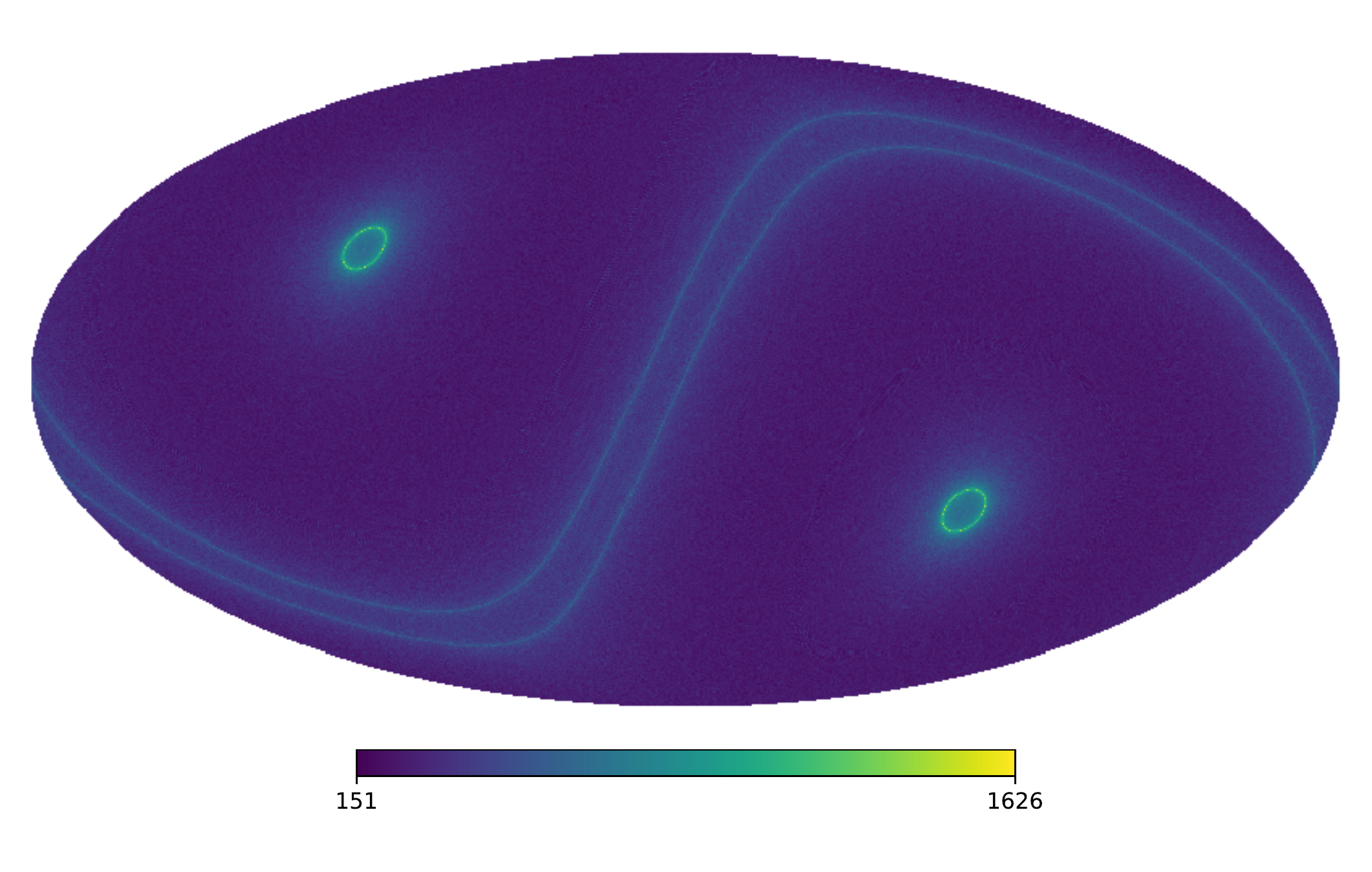}
    \caption{The hit map of the \textit{EPIC} scan strategy. This survey has been designed to maximise crossing angle coverage, and the hit map is well filled across the full sky with many observations at different orientations. Note that the galactic poles have been prioritised to aid the understanding of foregrounds and galactic science goals hence the higher number density in the hit map in those regimes}
    \label{fig:Hit Map}
\end{figure}

\begin{figure}[t!]
    \centering
    \includegraphics[width=1.\columnwidth]{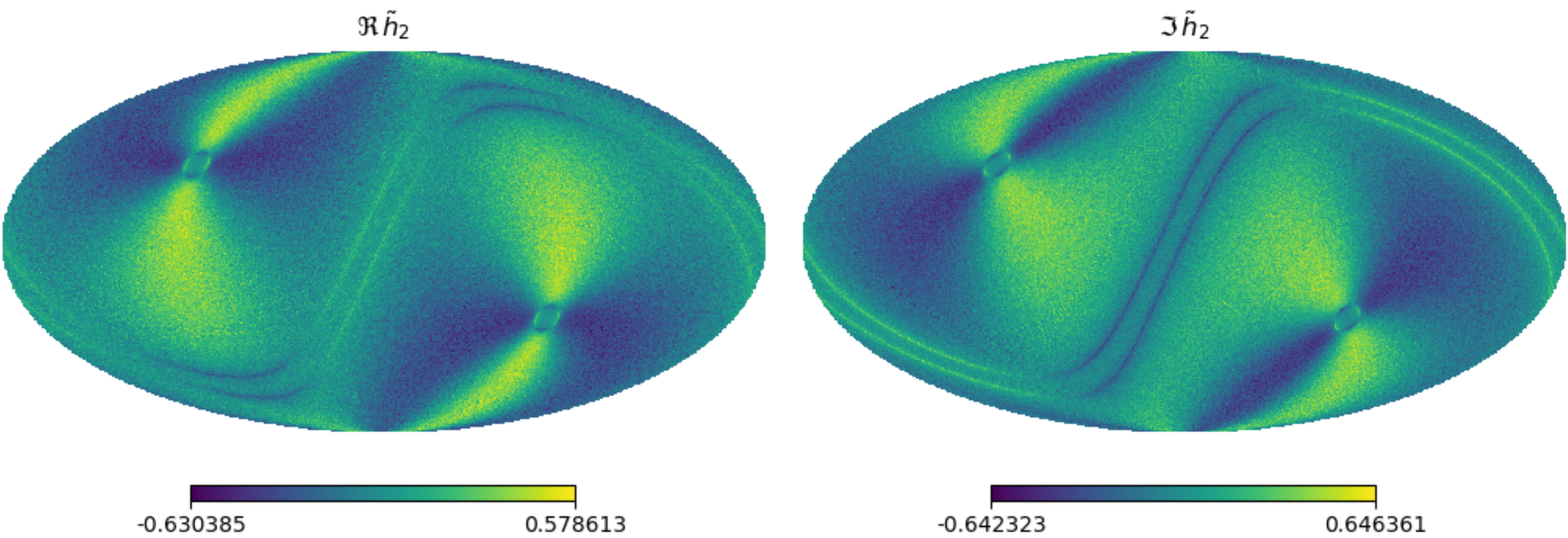}
    \caption{The real (left panel) and imaginary (right panel) parts of the $\tilde{h}_2$ field (defined in equation \ref{eqn:h_k}), which encodes the spin-2 part of the scan that turns the gain systematic into a spurious spin-2 signal. The $\tilde{h}_2$ field is dominated by its large scale features, which will result in a low $l$ dominated systematic.}
    \label{fig:Scan Maps}
\end{figure}

The scan strategy of \textit{EPIC} provides a relatively uniform distribution of hits and crossing angles which should reduce scan coupled differential systematics fairly well. The galactic poles are observed more frequently to aid in foreground analysis and galactic science goals, which results in some structure appearing in the survey fields as seen in Figure~\ref{fig:Hit Map}. The $\tilde{h}_{\pm 2}$ field (equation~\eqref{eqn:h_k}) encodes the spin-2 part of the scan that turns the gain systematic into a spurious spin-2 signal. This field is shown in Figure~\ref{fig:Scan Maps} for the \textit{EPIC} scan strategy. The $\tilde{h}_2$ is dominated by its large scale features, and this will result in a low $l$ dominated systematic.

\subsection{Simulation\label{sub:Simulation}}
We use a modified version of the code used in \cite{Wallis2016}. The input to the time ordered data (TOD) simulation code consists of maps of the CMB $\tilde{T}$, $\tilde{Q}$ and $\tilde{U}$ fields which are generated using the {SYNFAST} routine of the {HEALPIX} package \citep{2005ApJ...622..759G}. The input CMB power spectra were created in CAMB using a six parameter $\Lambda$CDM cosmological model, specified in Table~\ref{tab:simulation inputs} \cite{Lewis:1999bs}.  For the simulations including noise, we include a Gaussian beam post process. A white noise is applied to the data at map level, where a noise is added to each pixel of the level $w_{TT}^{-1} = 2.7\mu \text{K arcmin}$ \citep{2019JLTP..194..443H}.
\begin{table}[h!]
    \centering
    \begin{tabular}{cl}
    \hline\hline
        &Simulation Inputs\\
        \hline\hline
         Cosmology & $H_{0}$ = 67.4\\
         &$\Omega_{b}h^{2} = 0.022$\\
         &$\Omega_{c}h^{2} = 0.120$\\
         &$\tau = 0.06$\\
         &$n_{s} = 0.97$\\
         &$10^{9}A_{s} = 2.2$\\
         &$r = 0.001$\\
         \hline
         Map-making &$N_{\rm side} = 2048$\\
         &$\ell_{\rm max} = 3000$\\
         &$|\delta g_{1}| = 0.01$\\
         &$|\delta g_{2}| = 0.01$\\
         \hline
    \end{tabular}
    \caption{The fiducial cosmological parameters and the map-making inputs for the TOD simulations. The simulated maps are smoothed by a Gaussian beam and noise is added per pixel.}
    \label{tab:simulation inputs}
\end{table}

\begin{figure}[ht!]
    \centering
    \includegraphics[width=0.9\columnwidth]{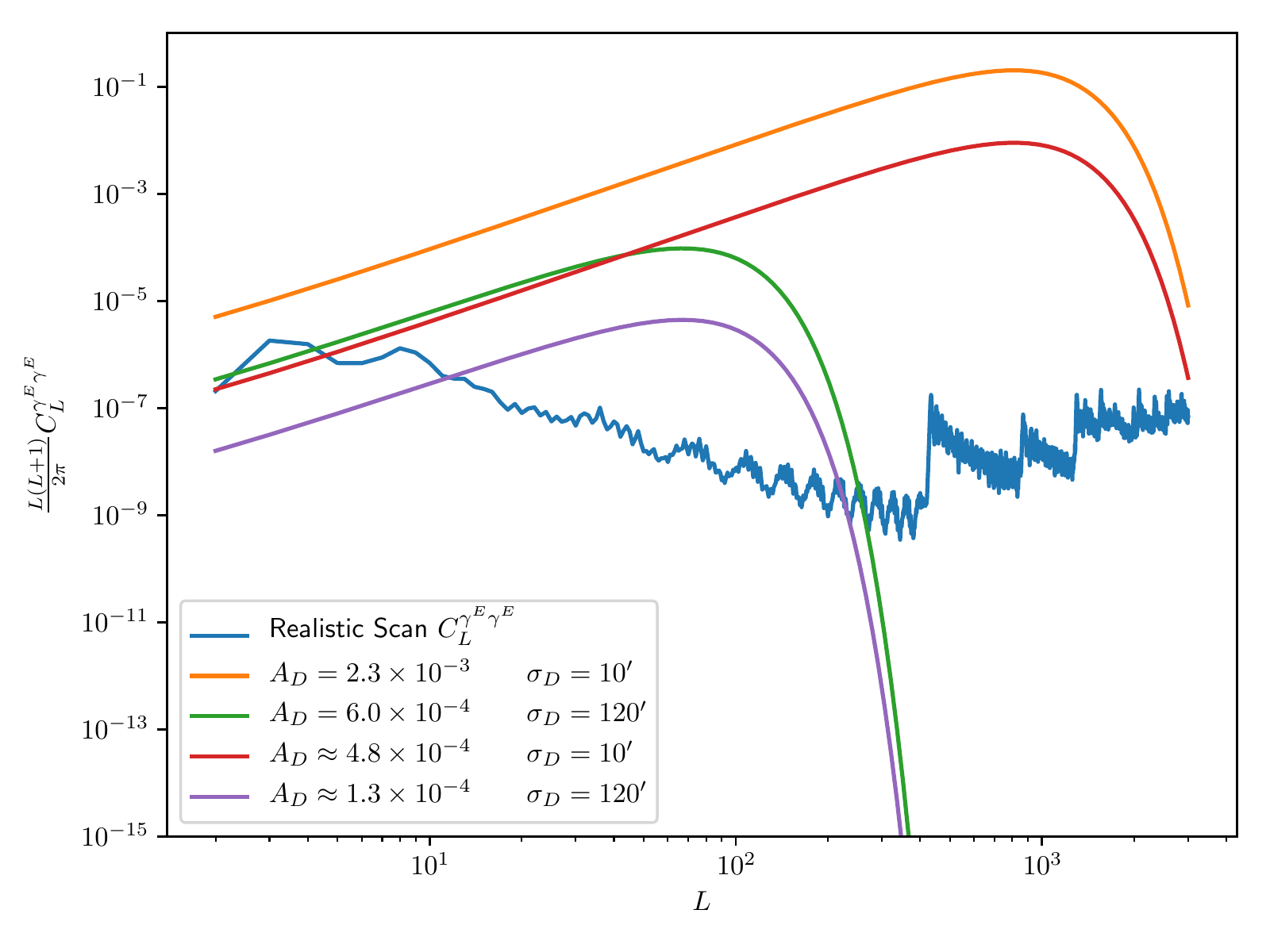}
    \caption{The E-mode spectrum of the ${}_{\pm}\gamma(\uvec{}n)$ distortion fields, the orange and green lines show the spectra plotted for the $A_{D}^{\text{Max}}$ amplitudes of \citep{2010PhRvD..81f3512Y} for coherences scales of $10^{\prime}$ and $120^{\prime}$ respectively, the red and purple lines show the spectra plotted for the $A_{D}^{\text{Min}}$ amplitudes of \citep{2010PhRvD..81f3512Y} for the CMBPol satellite estimate corresponding to r=0.005 for coherences scales of $10^{\prime}$ and $120^{\prime}$ respectively. The blue line shows the realistic spectrum calculated from the a combination of the systematic and the \emph{EPIC} scan strategy that we adopt for our simulations. It is clear here that the approximation used for the distortion power spectra does not describe the realistic case well at higher multipoles. It overestimates the power at high multipoles in comparison to the realistic case for small coherence scales. Additionally the sharp drop off in power after the peak is not consistent with the realistic spectra as is evident for all sets of curves showing the approximation.}
    \label{fig:Realistic Gamma Comparison}    
\end{figure}
%

The simulation computes values for each of the four detectors for each hit as described in equation \eqref{eqn:singledetect}, using HEALPIX interpolation to observe the input sky maps at the appropriate location, and using the corresponding parallactic angle $\psi$ for that hit, generated from the \textit{EPIC} scan strategy. As described in section \ref{sec:T to P Sims}, the $\psi$ values are offset by $90^\circ$ for the two detectors within a pair, and by $45^\circ$ between the two pairs of detectors. The differential gain systematic is added for each pair of detectors by increasing the signal by a factor $(1-\delta g_{i})$ in the second detector $d^B_i$ in each pair, where we use $|\delta g_{1}| = |\delta g_{2}| = 0.01$ for the simulations in this work \citep{2020MNRAS.491.1960T}. This level of systematic is indicative of the differential gain seen in recent CMB surveys \citep[e.g.][]{2015ApJ...814..110B,1403.2369}, and corresponds to an $r$ bias of $\sim30\sigma$ .

Maps are made from the time streams according to a simple binning map making technique,
\begin{equation}
\begin{pmatrix}Q\\U\end{pmatrix}
=
\begin{pmatrix}\langle \cos^{2}(2\psi_{i}) \rangle&\langle \cos(2\psi_{i})\sin(2\psi_{i}) \rangle\\\langle \sin(2\psi_{i})\cos(2\psi_{i}) \rangle&\langle \sin^{2}(2\psi_{i}) \rangle\end{pmatrix}^{-1}
\begin{pmatrix}\langle d_{i}\cos(2\psi_{i}) \rangle\\\langle d_{i}\sin(2\psi_{i}) \rangle\end{pmatrix}\;,
\label{eq:2x2 map making}
\end{equation}
where the angle brackets $\langle \rangle$ denote an average over the measurements in a pixel, and the $d_j$ here correspond to the detector measurements (i.e. the sum of the timestreams from the two differenced pairs) each with its an associated angle $\psi_j$. 

\subsection{Realistic Systematic Spectra}
\label{sec:Realistic Scan}
In previous studies \citep[e.g.][]{2003PhRvD..67d3004H,2010PhRvD..81f3512Y} the distortion fields of equation \ref{eqn:Distortion Fields} have been assumed to be statistically isotropic and Gaussian, and defined by power spectra of the form
\begin{equation}
    C_{l}^{DD} = A_{D}^{2} e^{-l(l+1)\sigma_{D}^{2}/8\ln{2}}\;,
    \label{eqn:Syst Cl Approx}
\end{equation}
where $A_D$ represents the root mean squared of the distortion field, and $\sigma_{D}$ represents a coherence scale beyond which the distortion power spectrum becomes white noise. We show a comparison between this spectrum and the realistic spectrum from our simulations in figure \ref{fig:Realistic Gamma Comparison}. There are significant differences apparent between the approximation of equation \ref{eqn:Syst Cl Approx} for the power spectra describing the distortion fields used in previous literature, and the realistic distortion field that is derived from the more realistic simulation used in our study.

Although the spectrum generated from equation \ref{eqn:Syst Cl Approx} results in most of the power being at low multipoles similar to the realistic spectra, it does not capture the high $l$ nature of the realistic distortion fields. The realistic distortion fields have an initial much sharper drop off, before levelling out, compared to the more gradual drop off of the approximation. In figure \ref{fig:Realistic Gamma Comparison} we show that the two extreme coherence scales used in \citep{2010PhRvD..81f3512Y} both suffer from the same issue that they accrue too much power at high $l$ compared to the realistic spectra, hence the results found will be biased by this. In our analysis in Section~\ref{sec:reconstruction} we reconstruct the distortion power spectrum up to $L=800$ and, as can be seen in figure \ref{fig:Realistic Gamma Comparison}, the shapes of the realistic and approximate spectra differ significantly for this range of multipoles.
This shows the advantage of carrying out detailed case studies on individual distortions.
\par 
We note  that the smallest  $A_D^{\rm min} \approx 
1.3\times10^{-4}$ for $\sigma_{D}=
120^{\prime}$ quoted from \cite{2010PhRvD..81f3512Y} corresponds to a gain mismatch of $
6.4\times10^{-4}~(0.064\%)$. This is significantly smaller than a realistic gain mismatch of 0.01 (1\%), in the analysis in \cite{2010PhRvD..81f3512Y} it was found that the QE technique would still be effective for this small value of the systematic.
\section{Reconstructing and removing the temperature to polarization leakage\label{sec:reconstruction}}
Here we discuss the results of employing the statistical analysis methods developed in Section~\ref{sec:qe} and Section~\ref{sec:iter_clean} to contaminated CMB maps  simulated as in Section~\ref{sec:T to P Sims}. Note that we include lensing effects only at the power spectrum level, implying that the off-diagonal elements sourced by weak lensing are not included in our simulations. This is not expected to influence our inference of the reconstruction and removal of the T to P leakage systematic, owing to different spins associated with the two effects\footnote{Note that this is not generally true for other instrument systematics. As an example a differential pointing systematic directly couples to the weak lensing effect and in a analogous study the weak lensing induced correlations cannot be ignored.}.

In all our analyses we iterate over the following steps until convergence:
\begin{itemize}
\item Reconstruct map of systematics given some input $[T,Q,U]$ maps using the QE algorithm. In the first iteration the inputs correspond to the observed maps, while for the subsequent iterations these correspond to the contamination cleaned maps.
\item Clean the input maps using the reconstructed $\gamma$ maps following the optimal cleaning procedure discussed in Section~\ref{sec:cleaning scheme}.
\end{itemize}
Here we reemphasize that the cleaning analysis is agnostic to details of the particular systematic, as evident by the fact that the cleaning procedure only works with the observed maps as inputs.  All QE evaluations required in the blind cleaning process are carried out assuming the parameter settings summarized in Table~\ref{tab:cleaning inputs}. We make forecasts for the reconstruction noise and the $C_{l}^{BB}$ spectrum that one expects to recover from the iteratively cleaned maps, following procedures outlined in Section~\ref{sec:iter_forecast}. We use these forecasted power spectra as benchmarks for our blind cleaning analysis.

We present the results of this analysis on two different set of simulations, the ideal case and the more realistic, non-ideal case, in sections \ref{sec:results_ideal} and \ref{sec:results_litebird} respectively. A discussion with particular emphasis on the measurement of tensor to scalar ratio $r$ is presented in Section~\ref{sec:r}.

\begin{table}[t]
    \centering
    \begin{tabular}{|c|c|c|}
        \hline
         Parameter & Ideal & Non-ideal\\
         \hline
         $l_{\rm max}$ & $1400$ & $1400$\\
         $L_{\rm max}$ & $800$ & $800$\\
         $N_{\text{side}}$ & $1024$ & $1024$\\
         $w^{-1}_{TT}$ & $0$ & $2.7 \,\mu\text{K}\,\text{arcmin}$\\
         $\theta_{FWHM}$ & $0$ & $30^{\prime}$\\
         $r$ & $10^{-3}$& $10^{-3}$\\
         $A_{\rm lens}$ & $0$& $1$\\
         \hline
    \end{tabular}
    \caption{This table summarizes the QE parameter settings used in all our analyses and also the simulation settings for the two different sets of simulations used in results presented in this section.}
    \label{tab:cleaning inputs}
\end{table}
%

\subsection{The cosmic variance limits of the blind cleaning algorithm} 
\label{sec:results_ideal}
Here we discuss the results derived from analyses on simulations which are ideal in the sense that they include no measurement noise \& beam smoothing and also do not include any lensing induced B-modes. These simulations allow us to probe the limitations of the blind cleaning algorithm in this extreme setting, in process highlighting how well this procedure could in principle work.
\begin{figure}[t]
    \centering
    \includegraphics[width=0.75\columnwidth]{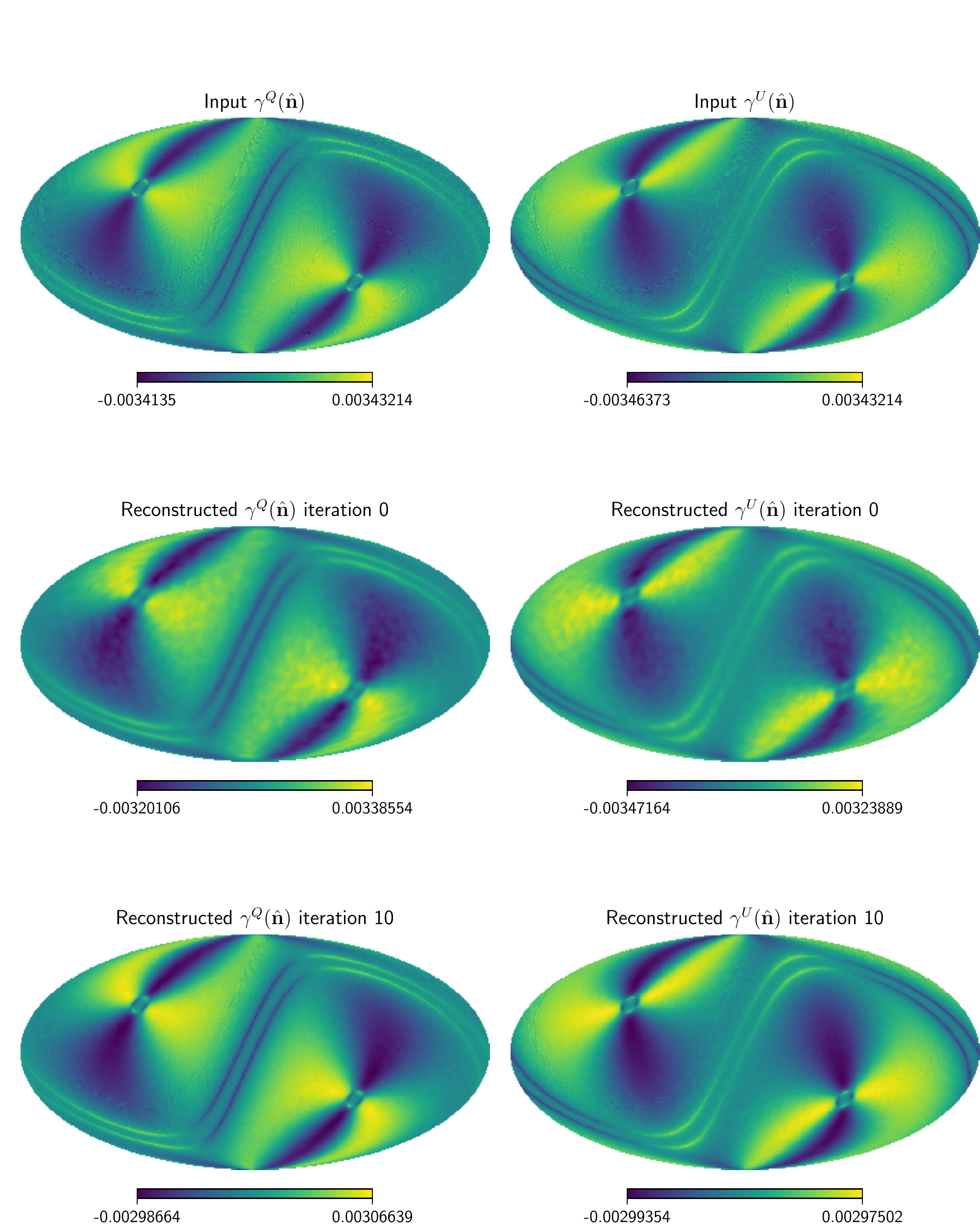}
    \caption{This figure depicts the $\gQ$ and $\gU$ systematic maps that mediate the T to P leakage. The top panels depicts the true systematic maps used to simulate the contaminated CMB maps. The middle panels shows the QE reconstruction of the $\gamma$ maps from the observed CMB maps for the 0th iteration. The bottom panels depict the reconstructed $\gamma$ maps after 10 iterations of cleaning and reconstructing of the systematic maps.}
    \label{fig:maps ideal 10th}
\end{figure}

\begin{figure}[t]
    \centering
    \begin{subfigure}[t]{0.49\textwidth}
    \includegraphics[width=1\columnwidth]{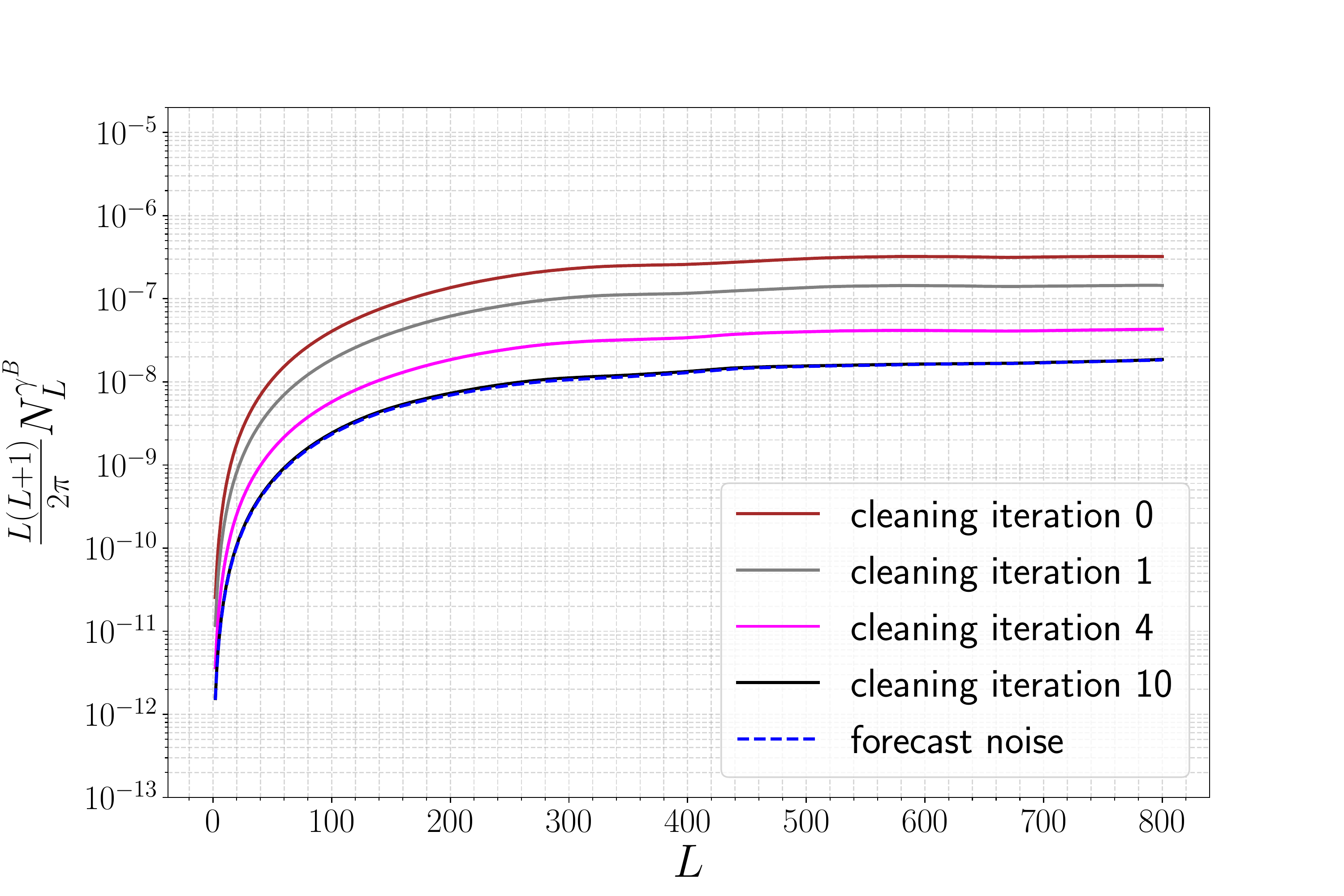}
    \end{subfigure}
    \begin{subfigure}[t]{0.49\textwidth}
    \includegraphics[width=1\columnwidth]{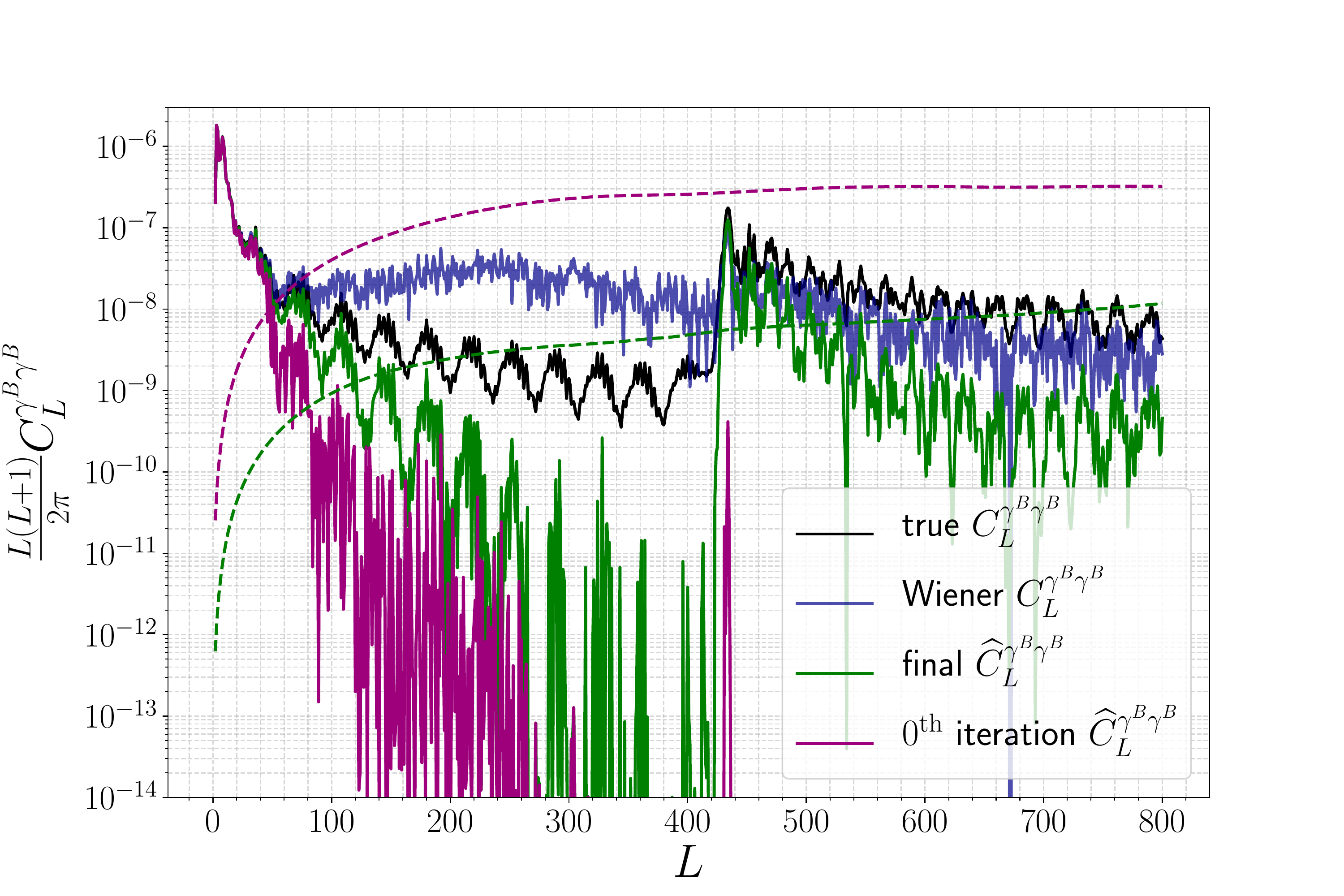}
    \end{subfigure}

    \caption{The the left panel shows the evolution of the reconstruction noise for different iterations of the algorithm. Note that the reconstruction noise reduces with iterations and approaches that predicted using the forecasting procedure. The right panel shows the power spectra of the true and reconstructed and filtered $\gamma$ maps. The corresponding reconstruction noise curves are plotted for reference. Note that the Wiener filtered, $\hat{C}^{\gamma^{B}\gamma^{B}, {\rm WF}}_{L}$, spectrum is most reliably recovered for modes where $C^{\gamma^{B}\gamma^{B}}_{L}$ is much greater than $N^{\gamma^{B}\gamma^{B}}_{L}$. Similar results found for $\gamma^E$, not show here for brevity.}
    \label{fig:nonoisenobeamreconnoise}
\end{figure}

\begin{figure}[t]
    \centering
    \begin{subfigure}[t]{0.49\textwidth}
    \includegraphics[width=1\columnwidth]{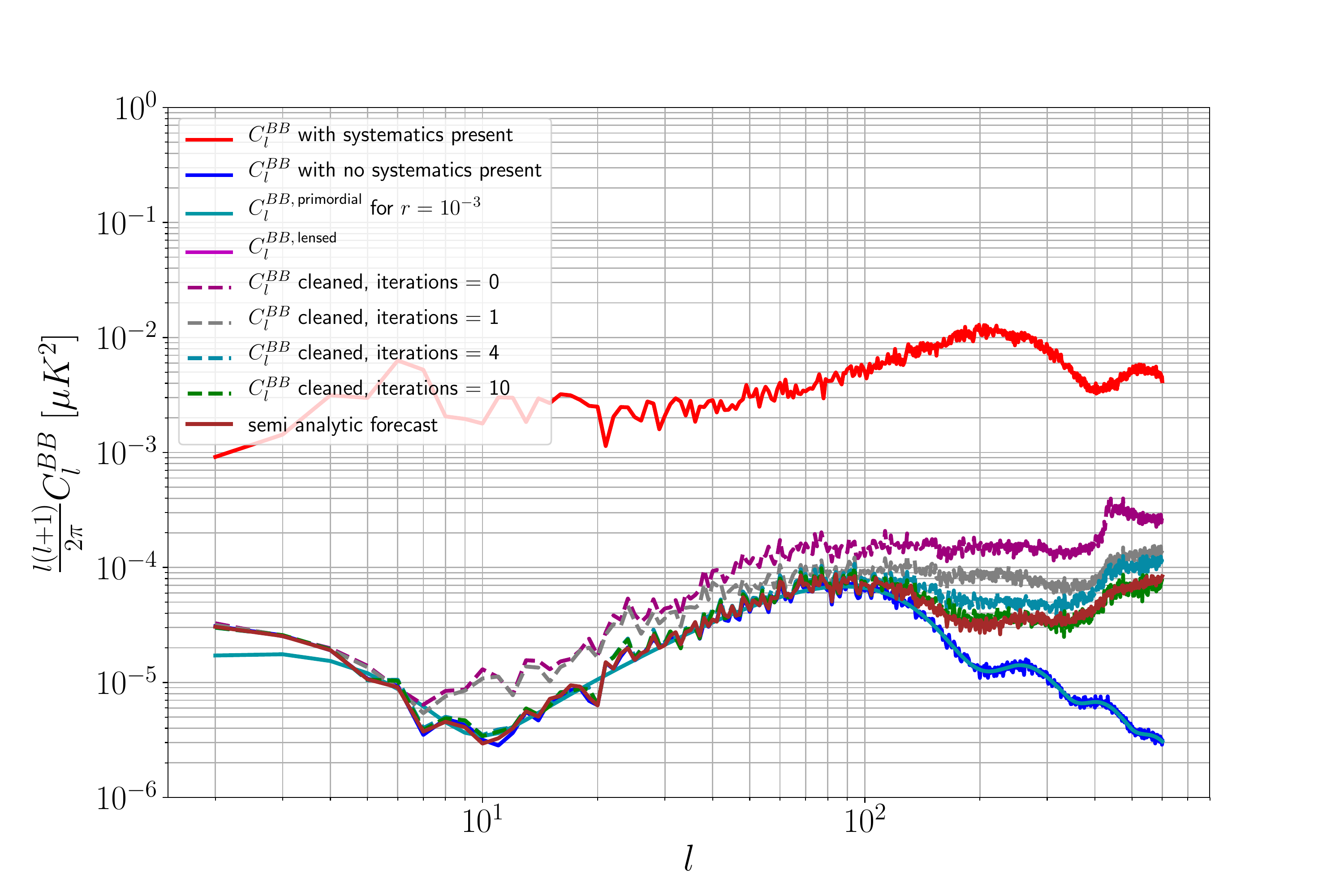}
    \end{subfigure}
    \begin{subfigure}[t]{0.49\textwidth}
     \includegraphics[width=1\columnwidth]{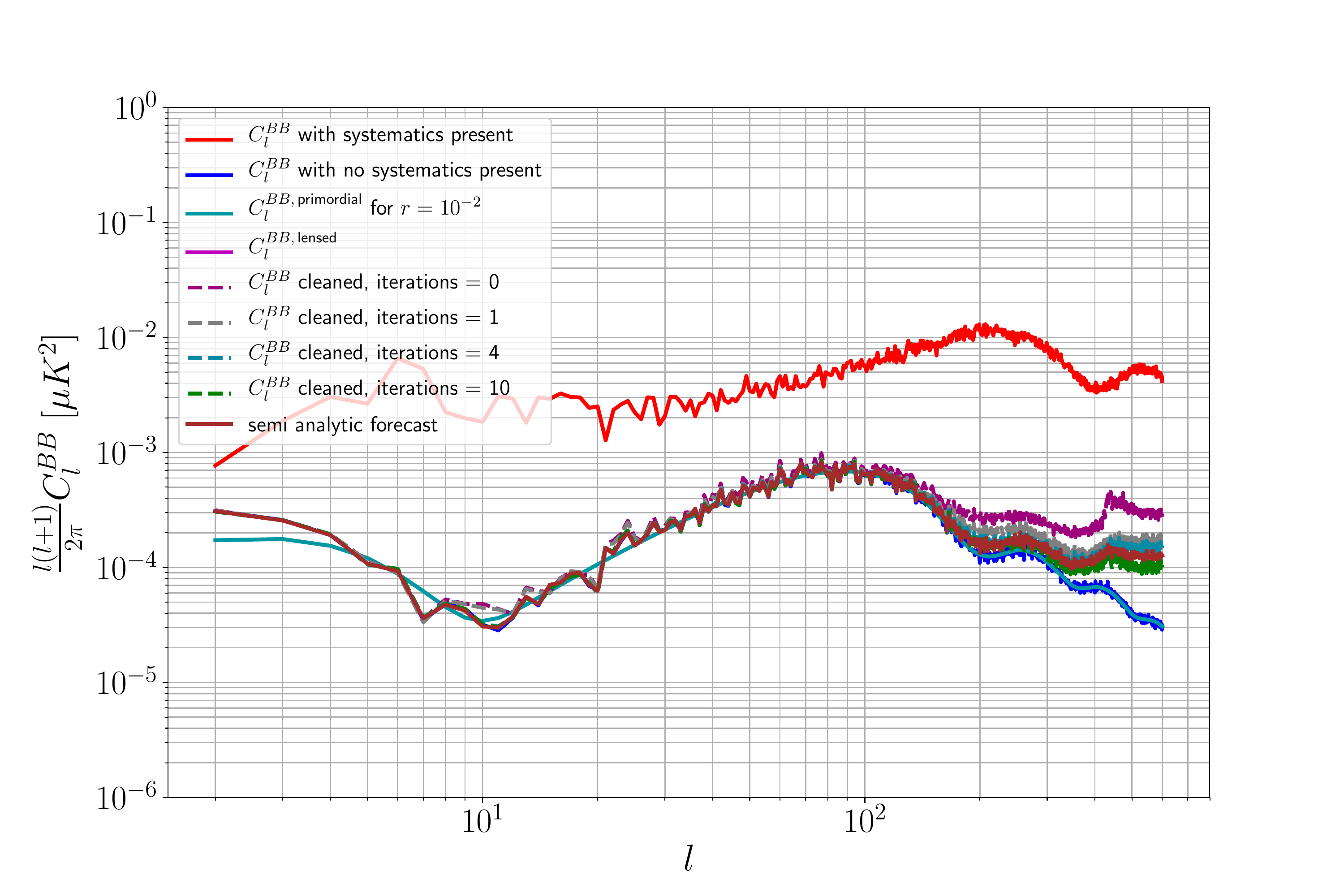}
    \end{subfigure}
    \caption{This figure depicts the B-mode power spectrum corresponding to the systematic ridden maps, the true cosmological primordial signal and the evolution of the estimated power spectrum across different iterations of the cleaning algorithm. The left and right panels show the cleaning for $r=10^{-3}$ and $r=10^{-2}$ respectively.  Also shown is the prediction for the B-mode power spectrum expected post cleaning evaluated using the forecasting procedure.}
    \label{fig:nonoisenobeamboth}
\end{figure}

The reconstructed $\gamma$ maps at some example iterations of the analysis are depicted in Fig.~\ref{fig:maps ideal 10th}. Here we note that the systematics reconstructed from the original observed maps are quite noisy as inferred by comparing the top and middle panels of Fig.~\ref{fig:maps ideal 10th}. This observation is better quantified in Fig.~\ref{fig:nonoisenobeamreconnoise}, where by inspecting the reconstruction noise for the $0^{\rm th}$ iteration and comparing it to the true $\gamma$ power spectrum, it is clear that only $L \lesssim 30$ multipoles of the $\gamma$ map can be reliably recovered. The reconstruction noise being high is due to the excess B-mode power sourced by the systematics in the observed CMB maps as seen in Fig~\ref{fig:nonoisenobeamboth}. 

We now use these reconstructed $\gamma$ maps together with the observed temperature anisotropy map to remove part of the contamination, sourced by modes in the $\gamma$ maps that have been reliably recovered. This procedure involves using the high SNR modes of the temperature and $\gamma$ maps, the formal details of which are discussed in Section~\ref{sec:cleaning scheme}. In the case of these idealized simulations, we find that the conventional Wiener filtering scheme causes the iterative scheme to diverge after few initial iterations. We suspect this behaviour arises from the fact that the Wiener filter does not sufficiently suppress modes that have a noisy recovery, which combined with the fact that $T>>B$, leads to a faulty cleaning of the polarization maps, in effect adding more power to the B-mode map as opposed to subtracting it. This eventually leads to run away behaviour. We deal with this issue by employing the Gaussian filtering scheme instead (see Section~\ref{sec:cleaning scheme} for details) which mitigates this issue by imposing a stronger suppression of the noisy modes, leading to more stable and convergent results. The right panel of Fig.~\ref{fig:nonoisenobeamreconnoise} depicts the power spectrum of the $\gamma^B$ maps after the Gaussian and Weiner filters are applied to them, where notably the Weiner filtered maps retain a lot of power from the unreliably recovered modes, as opposed to the Gaussian filtered map where these noisy modes are more heavily suppressed.

After the first cleaning (i.e. cleaning iteration 0), the B-mode power spectrum reduces compared to the spectrum estimated from the observed B-mode map as seen in Fig~\ref{fig:nonoisenobeamboth}. This results in the reconstruction noise of the QE to reduce as can be understood by comparing the curves corresponding to "iteration 0" and "iteration 1" in Fig.~\ref{fig:nonoisenobeamreconnoise}. This reduction in the reconstruction noise, facilitates the recovery of modes in the $\gamma$ map that were dominated by the reconstruction noise in the previous iteration. These newly recovered modes of the systematic map are then fed to the cleaning algorithm to further remove the contamination from the polarization maps. This whole process is repeated until we observe no further improvements in either the reconstruction noise and/or the $C_{\ell}^{BB}$ spectrum.
\par
On repeating this procedure we see that the contamination in the CMB polarization maps is progressively removed as indicated by the systematic reduction in the amplitude of $C_l^{BB}$ amplitude in Fig.~\ref{fig:nonoisenobeamboth}. Note that initial iterations show relatively big reductions in power, with subsequent iterations resulting in more subtle improvements and the final few iterations show no appreciable updates to the spectrum. The performance of the cleaning process improves if the amplitude of the uncontaminated $B$-mode spectrum increases. For the ideal case this translates to improved performance for larger values of $r$. We note that even in this perfect setting of no instrument and lensing noise, the recovered B-mode spectrum is not perfectly cleaned. This can be interpreted as an intrinsic limitation to how well the cleaning can in principle perform. Nonetheless, the proposed blind cleaning procedure enables robust removal of contamination power that is roughly two orders of magnitude larger than the injected signal, and yields an unbiased recovery of the true signal at most multipoles. 
This systematic reduction in the B-mode power is only possible due to the simultaneously reduction in the reconstruction noise (sourced by reduction in $C_{\ell}^{BB}$) as seen in Fig.~\ref{fig:nonoisenobeamreconnoise}, which results in robust recovery of the higher multipole of the $\gamma$ maps (which in turn facilitates more cleaning of the polarization maps). To contrast the effect of iterative cleaning note that while the "iteration 0" only allowed for recovery of the modes $L \lesssim 30$, the reconstruction noise associated with the final iteration of cleaning allows robust recovery of modes up to $L \simeq 800$ as can be seen in right panel of Fig.~\ref{fig:nonoisenobeamreconnoise}. This stark improvement in the recovery of high $L$ modes of the $\gamma$ maps can be better appreciated by simultaneously comparing the recovered total $\gamma$ maps shown in the bottom panels of Fig.~\ref{fig:maps ideal 10th} to those depicted in the panels above. Note that the total systematic maps is recovered by adding together the filtered maps of systematics estimated at each iteration\footnote{The $\gamma$ maps recovered at each iteration do not include the modes that were in effect subtracted from the polarization data in the previous iteration.}. The input maps have a higher amplitude than those reconstructed which is primarily a consequence of our maps being filtered and the reconstruction being terminated at $L_{\rm max}=800$. \par 
Here, it is also important to appreciate the non-monotonic nature of the true $C_{L}^{\gamma \gamma}$ which features a prominent jump in power at $L \sim 400$. This is a consequence of using realistic scan maps in our simulations. There is a corresponding feature in the cleaned $B$-mode spectra in Figure~\ref{fig:nonoisenobeamboth}. This feature is present because not all modes below $L \lesssim 800$ are reconstructed, as some intermediate modes which are dominated by reconstruction noise are suppressed. We reiterate that this would not have been observed in studies using the approximate spectra, \citep[e.g.][]{2003PhRvD..67d3004H,2010PhRvD..81f3512Y}, generated using equation~\eqref{eqn:Syst Cl Approx} because of the difference in shape between the realistic and approximate spectra. It is necessary to carry out detailed case studies systematics in order to observe these important details. Unlike in weak lensing studies, for instrument systematics it is not possible to make a generic forecasts as was done in \citep{2010PhRvD..81f3512Y}. 

Finally, we note that the spectrum converges to the prediction from our forecasting procedure. This is true both for $C_l^{BB}$ as well as $N_{L}^{\gamma \gamma}$ as seen in Fig.~\ref{fig:nonoisenobeamboth} and Fig.~\ref{fig:nonoisenobeamreconnoise}  respectively. It is important to note this near consistency for two reasons, (i) it serves as a validation of our blind cleaning algorithm (ii) the actual analysis is performed using the Gaussian filter, while our forecasting procedure continues to use Weiner filters, and the near equivalence of the two solutions suggests that the Gaussian filtering is close to optimal.

\subsection{Employing blind systematic cleaning for a non-ideal experiment} \label{sec:results_litebird}
\begin{figure}[t]
    \centering
    \begin{subfigure}[t]{0.49\textwidth}
    \includegraphics[width=1\columnwidth]{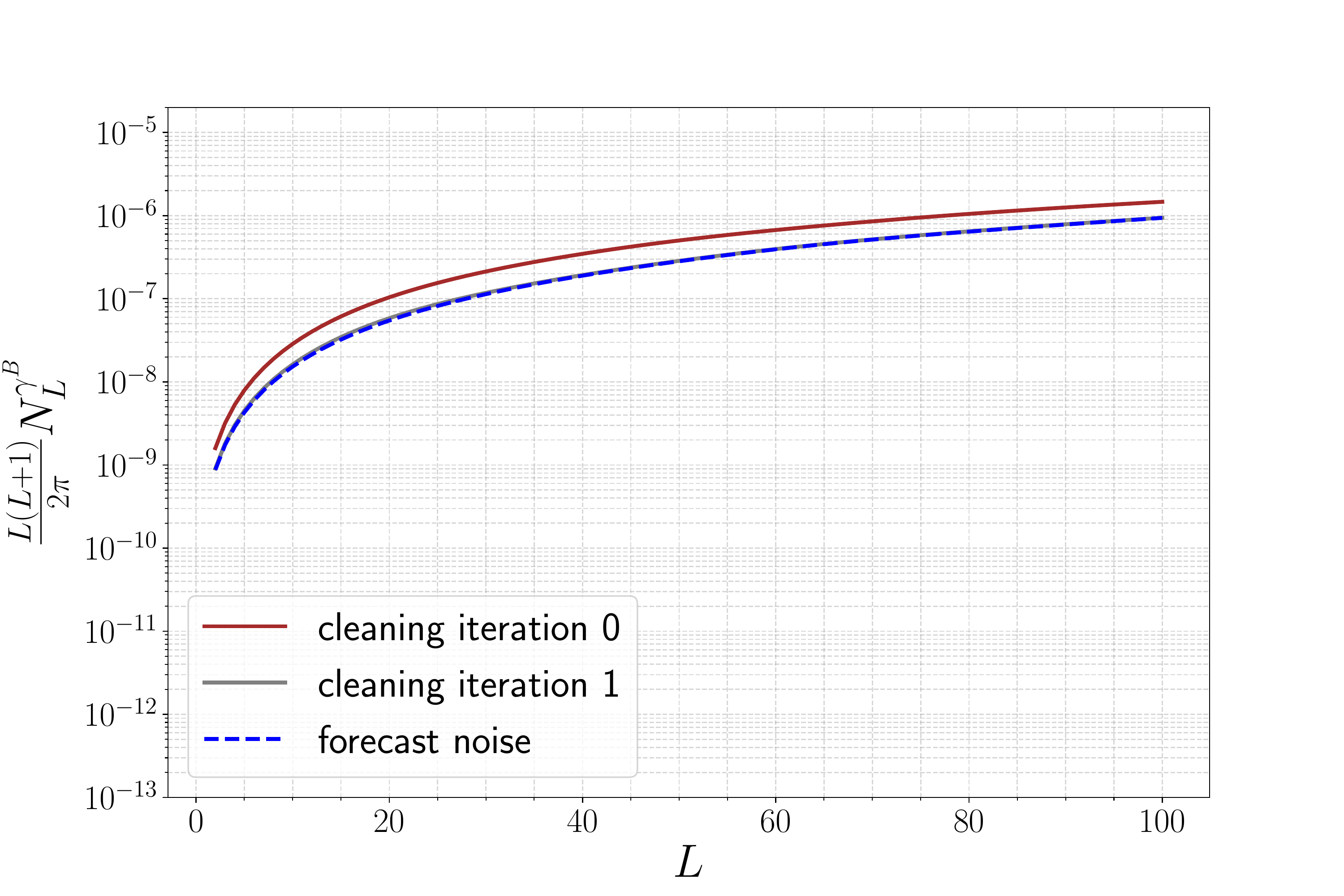}
    \end{subfigure}
    \begin{subfigure}[t]{0.49\textwidth}
    \includegraphics[width=1\columnwidth]{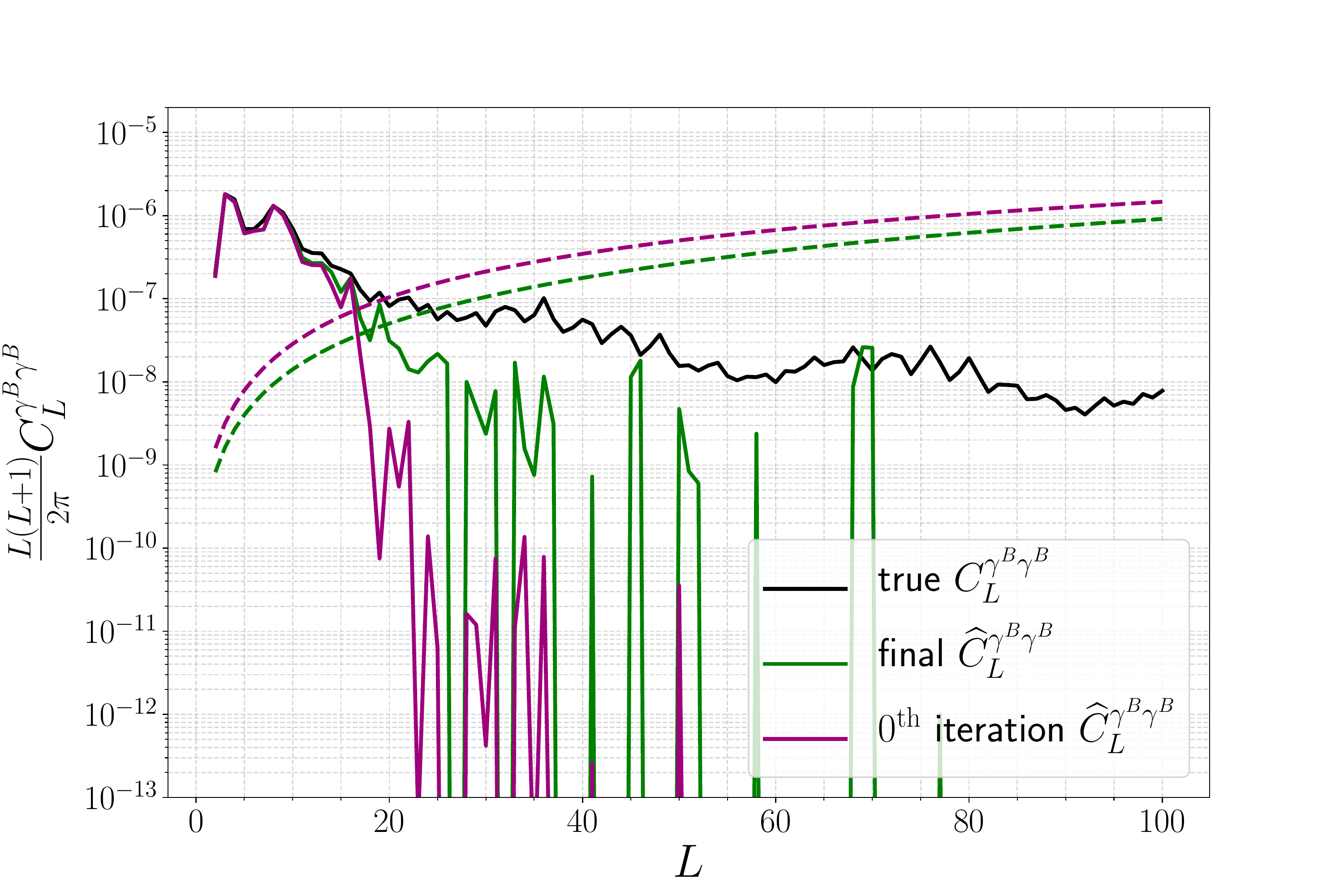}
    \end{subfigure}
    \caption{The left panel shows the reconstruction noise $N^{\gamma^{B}}_L$ for one iteration of the reconstruction and cleaning process. After a single iteration the cleaning reconstruction noise converges with the forecasted reconstruction noise. The right panel shows $\widehat{C}^{\gamma^{B}\gamma^{B}}_{L}$ for the reconstructed and filtered $\gamma$ maps before cleaning and after ten iterations  of cleaning (green) and the true $C^{\gamma^{B}\gamma^{B}}_{L}$ is also shown for comparison. Corresponding reconstruction noise shown for reference. }
    \label{fig:nb lensing recon}
\end{figure}
Here we discuss results simulations that incorporate weak lensing induced B-modes as well as the measurement noise and beam smoothing in the previously described more realistic, non-ideal case. The assumed measurement noise and beam are summarized in Table~\ref{tab:cleaning inputs} and correspond to the foreground cleaned Q/U maps that will result from linearly combining the multi-frequency measurements. Unlike in the previous section where these were ignored, here we treat the lensing $B$-modes as an important cosmological signal that we recover by carrying out the iterative de-contamination procedure.
\begin{figure}[t]
    \centering
    \includegraphics[width=0.75\columnwidth]{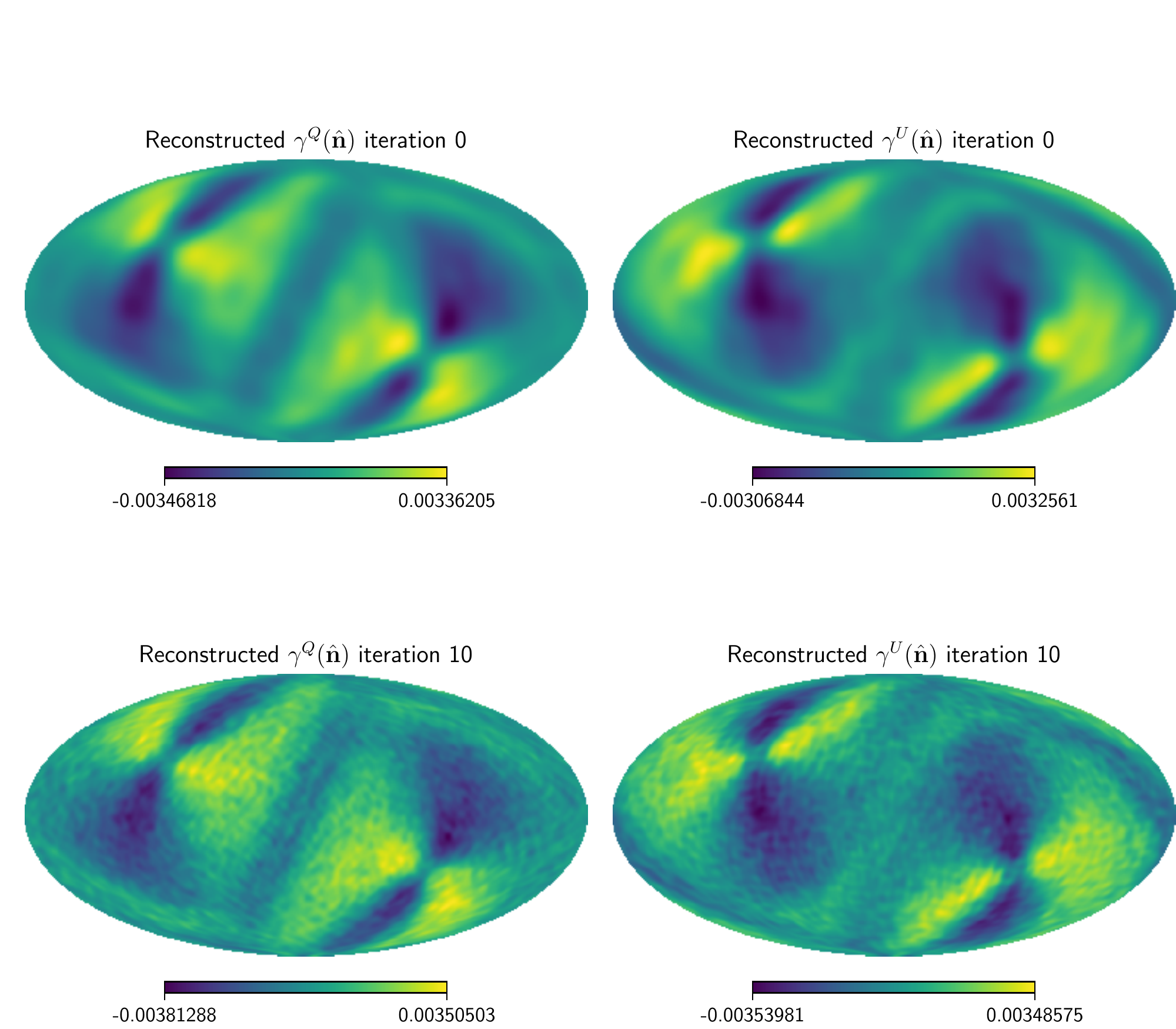}
    \caption{The top panels show the reconstructed filtered $\gamma$ maps recovered from simulated observations, while the bottom panels show those recovered after a few iterations of cleaning. Note that even in the presence of instrument and lensing noise the iterative procedure helps with extracting bits of information on the systematics.}
    \label{fig:maps lensing 10th}
\end{figure}
We carry out an analysis, identical to that described in the previous section, on these more realistic simulations.
The simulations used here primarily differ from those used in the previous section by inclusion of the relatively high noise in the observed maps due to inclusion of lensing and measurement noise. We note that for these relatively high noise simulations, the Wiener filtering schemes is stable and convergent, and the results are very similar to those found when using the Gaussian filtering. We however continue to present results derived from employing the Gaussian filtering scheme through the rest of the paper. The relatively high noise results in a higher QE reconstruction noise floor, which consequently limits the reliable reconstruction of the $\gamma$ maps to only the large angle modes $L \lesssim 20$, even after ten iterations of cleaning as seen in Fig.~\ref{fig:nb lensing recon}. This is even reflected in the total reconstructed $\gamma$ maps as seen in Fig.~\ref{fig:maps lensing 10th}. However note that the carrying out a number of iterative cleaning procedures does help in recovering some additional features in the reconstructed $\gamma$ maps, the sharpening of the features in the equatorial plane in the bottom left panel of Fig.~\ref{fig:maps lensing 10th} is particularly noticeable. As we will see in Section~\ref{sec:r} these subtle improvements in recovery of the systematic maps will play a crucial role in more robust removal of contaminations from the observed maps.

\begin{figure}[t]
    \centering
    \begin{subfigure}[t]{0.49\textwidth}
    \includegraphics[width=1\columnwidth]{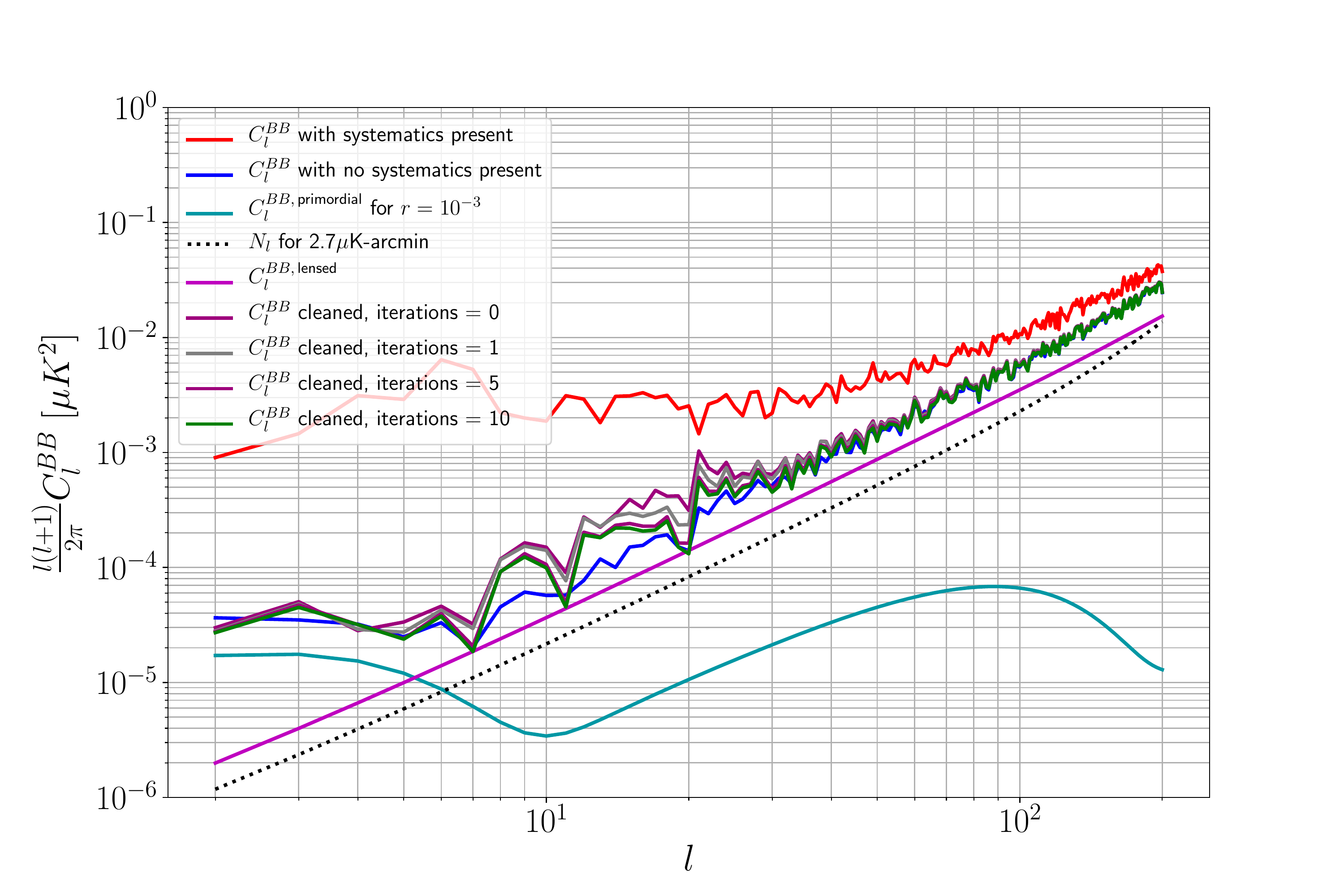}
    \end{subfigure}
    \begin{subfigure}[t]{0.49\textwidth}
    \includegraphics[width=1\columnwidth]{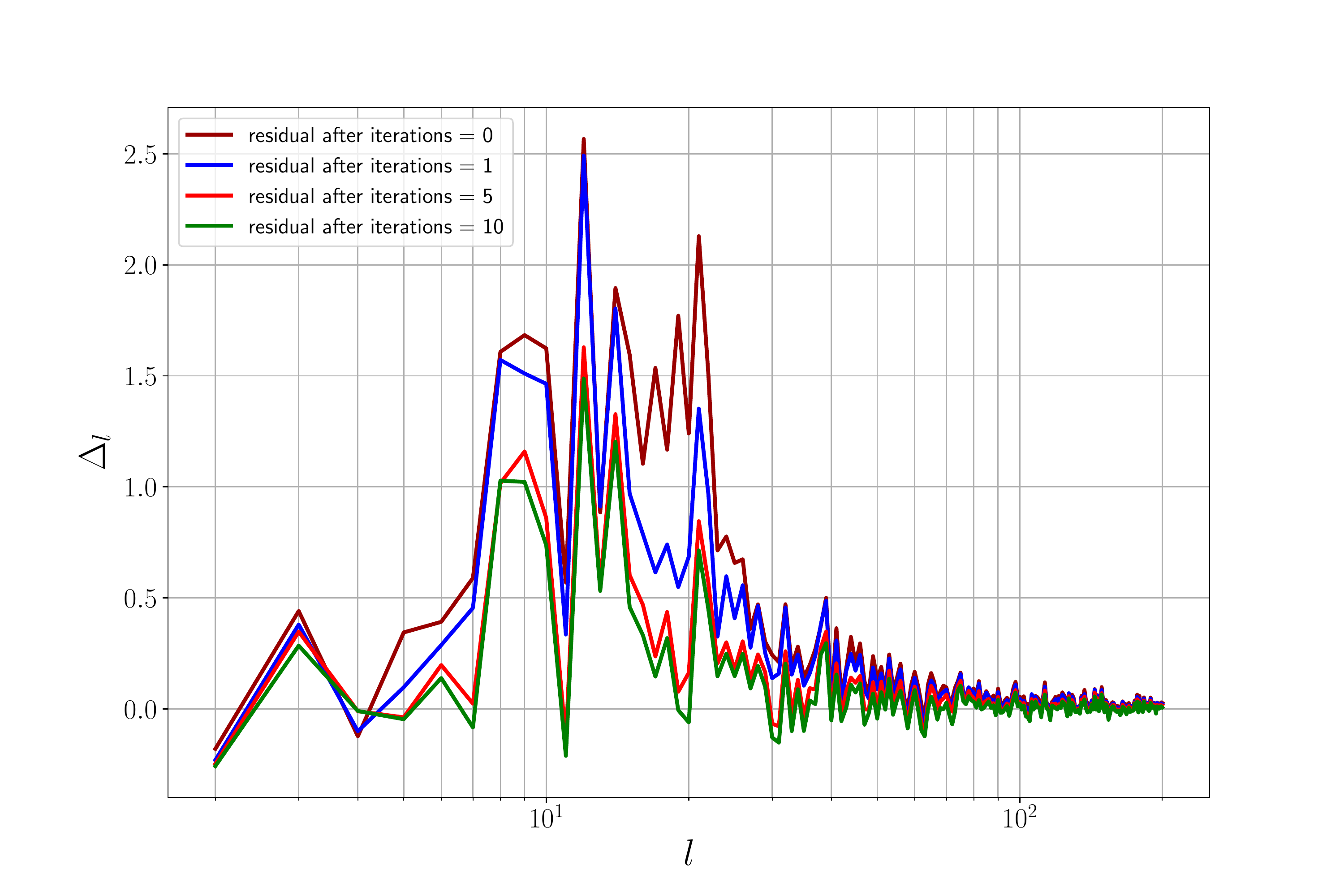}
    \end{subfigure}
    \caption{The left panel shows cleaned $B$-mode power spectra for 1 and 10 iterations of cleaning. 
    The right panel shows the evolution of the relative difference $\Delta_{l} = (C^{BB,\,\text{clean}}_{l}-C^{BB,\,\text{true}}_{l})/C^{BB,\,\text{true}}_{l}$ with iterations.}
    \label{fig:nb lensing cleaning}
\end{figure}

We now shift our attention to the evolution of $C_l^{BB}$ across the cleaning iterations. Carrying out higher iterations of cleaning does make small improvements to the convergence. These subtle but important improvements are highlighted in the right panel of Fig.~\ref{fig:nb lensing cleaning}, where the relative differences between the spectra derived from the cleaned maps at different iterations and true spectrum are depicted. In Section~\ref{sec:r} we will highlight the importance of these subtle corrections in the context of measurement of tensor to scalar ratio $r$.

\section{Recovering the tensor-to-scalar ratio\label{sec:r}}
In the previous section we demonstrated that blind systematic cleaning method proposed here can yield nearly un-biased recovery of the true CMB B-mode power spectrum. Upcoming experiments aim to recover $r \in [10^{-2},10^{-3}]$ \cite{Hazumi2019,Ade:2018sbj}. Here we demonstrate that the blind cleaning technique can yield nearly un-biased recovery of $r$. 
\par 
To demonstrate this we carry out a likelihood analysis for which we assume this specific form of the log-likelihood \cite{Hamimeche2008,Katayama_2011}, which accounts for the non-Gaussian nature of the power spectrum at low multipoles,
\begin{eqnarray}
    -2\ln\mathcal{L}(r) =&& \sum_{l}(2l+1)\Bigg[\frac{\widehat{C}^{BB}_{l}}{r\,C^{BB,\,GW}_{l}+C^{BB,\, \text{lens}}_{l} + N^{BB}_{l}} \\
    &&+ \ln\left(r\,C^{BB,\,GW}_{l}+A_{\rm lens} C^{BB,\, \text{lens}}_{l} + N^{BB}_{l}\right) - \frac{2l-1}{2l+1}\ln\left(\widehat{C}^{BB}_{l}\right) \Bigg]+\text{const.}\,, \nonumber
\end{eqnarray}
where $\widehat{C}^{BB}_{l}$ denotes the power spectrum estimated from the simulated data, corrected for the instrument beam, $C^{BB,\,GW}_{l}$ is the $B$-mode signal generated by primordial gravitational waves evaluated for $r=1$, $C^{BB,\,\text{lens}}_{l}$ denotes the lensing induced $B$-mode spectrum and $N^{BB}_{l}$ is the instrument noise power spectrum. We evaluate this likelihood analysis on power spectra derived from the cleaned maps at a number of different iterations and compare the estimated posteriors on $r$ to those derived from an analysis on a contamination free simulation. 

%
\begin{figure}[t!]
    \centering
    \begin{subfigure}[t]{0.9\textwidth}
    \includegraphics[width=1\linewidth]{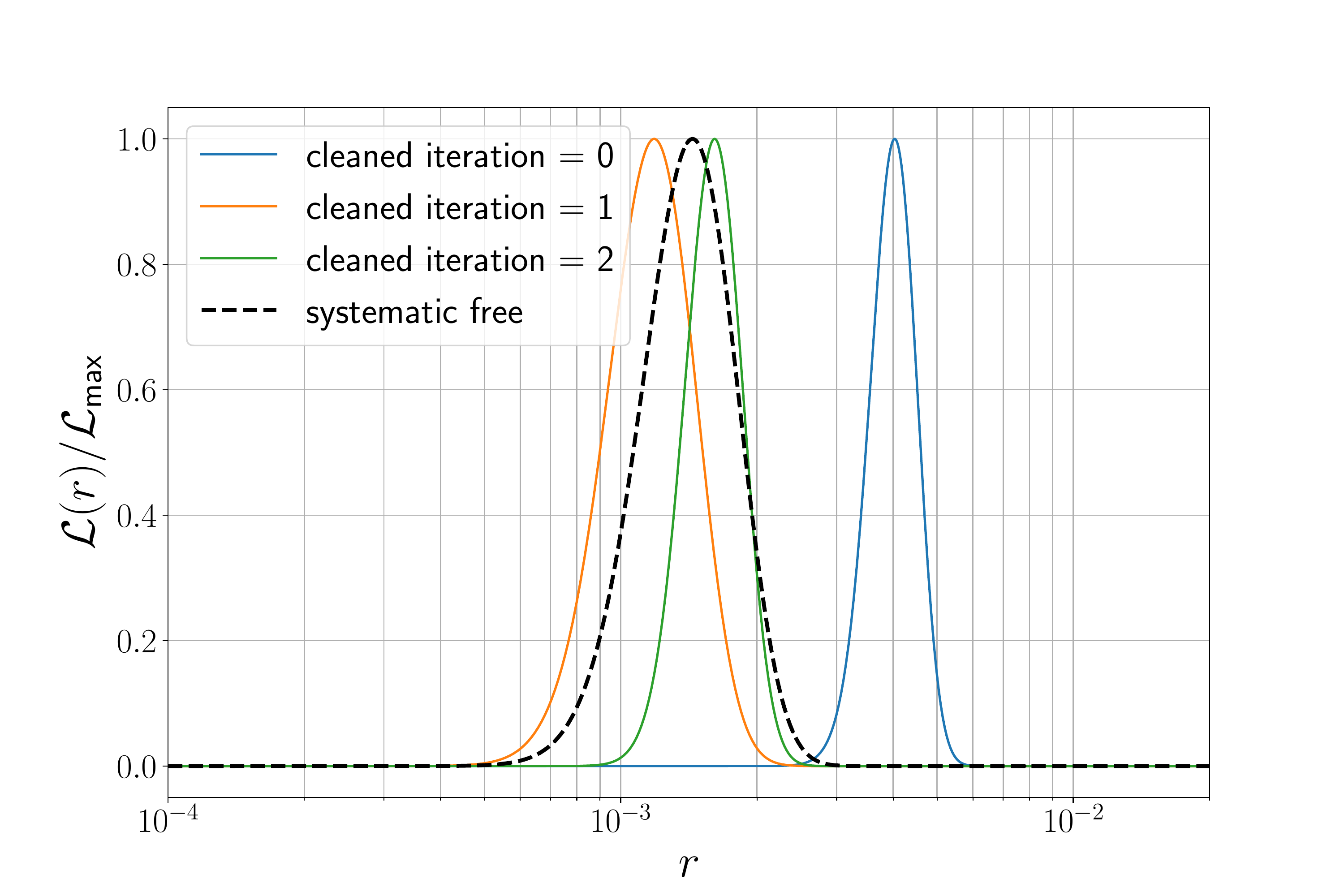}
    \end{subfigure}
    \caption{The evolution of the $r$-posterior across different cleaning iterations using the Wiener filter. The bias on $r$ reduces with each iteration and is remarkably consistent with the posterior derived from the systematic free simulation.}
    \label{fig:likelihood lensing}
\end{figure}
We begin by noting that the $r$ inferred from the contaminated $B$-mode simulations returns a highly biased measurement of $r\sim 10^{-1}$, off-set by two orders of from the true value. However, on repeatedly applying the iterative cleaning algorithm to it using either the Wiener or the Gaussian filter, the bias in the measurement of $r$ reduces. The reduction in $r$-bias is largest in the non-ideal setup when using Wiener filtering. As seen in Fig.~\ref{fig:likelihood lensing}, after 2 iterations we find that the measured value of $r$ is consistent with the true value to better than $1 \sigma$. This corresponds to a near perfect removal of a bias of order $\sim100$.Beyond the 2nd iteration of cleaning there is no further change to the likelihood.  Note that this cleaning also reduces the uncertainty on $r$ by a similar order of magnitude. With lower values of $A_{\text{lens}}$, corresponding to delensed maps, $r$ is still consistent with the true value. However, more iterations are required to achieve convergence after delensing, as the reconstruction noise floor is lowered. Delensing does not significantly improve the level of cleaning that it is possible to achieve with the more realistic, non-ideal case. A joint study of delensing and systematics cleaning could become more relevant in the case of higher sensitivity experiments such as PICO \cite{hanany2019pico}.
\begin{table}[t]
    \centering
    \setlength{\tabcolsep}{10pt}
    \begin{tabular}{|ccccc|}
    \hline
        
%
%
            &iteration& $r$ & $r_{-}$ & $r_{+}$  \\\hline
          Contaminated & - & $113$ & $111$ & $115$\\\hline
        
         &0 & $4.03$ & $3.58$ & $4.53$\\
          Cleaned   & 1 & $1.19$ & $0.972$ & $1.5$\\
         &2& $1.61$ & $1.41$ & $1.88$\\
          &3& $1.61$ & $1.41$ & $1.88$\\
          \hline
           
           True & - & $1.44$& $1.16$ & $1.86$\\\hline  
    \end{tabular}
    \caption{This table presents the central value of $r$, and  68\% CI upper and lower bounds in units of $10^{-3}$. These results assume no de-lensing i.e. $A_{\text{lens}}=1$.}
    \label{tab:distortion r}
\end{table}

\section{Conclusions}
\label{sec:conclusions}

Systematic effects originating in the instrument pose a major challenge for upcoming CMB experiments seeking to measure the primordial CMB $B$-mode of polarization. Many of the existing techniques for mitigating these effects rely on complex instrument modeling and detailed knowledge of the instrument design.
We have presented a detailed case study, implementing a QE approach to carry out cleaning of the CMB $B$-mode without detailed prior knowledge of the instrument. We have shown that this QE technique can successfully remove a T to P leakage sourced by a differential detector gain systematic, resulting in a near optimum recovery of the primordial $B$-mode and the reduction of the bias on the tensor-to-scalar ratio by $\sim2$ orders of magnitude.  Our robust implementation builds on the previous work by carrying out systematic recovery and map correction on a full TOD simulation including the effects of a realistic satellite scan strategy, and by the use of newly-derived efficient full-sky estimators. In our recovery and map correction we use a novel Gaussian filter which we find to be an effective alternative in cases where the Wiener filter caused the map correction to fail.
\par
Our case study involved two scenarios.  The first scenario, with no noise, beam or lensing, was used to illustrate the absolute limit to the cleaning process in an ideal world when there are no complications. The second scenario provides a more realistic, non-ideal example of the cleaning for a contemporary CMB experiment by using realistic levels of noise and beam comparable to those expected for the \emph{LiteBIRD} instrument \cite{Hazumi2019}. It was necessary to test the iterative cleaning scheme used in our map correction. We carried out this testing using a semi-analytical forecast for the ideal and realistic cleaning. Our cleaning was successful as it was found to be consistent with the forecast
\par
We used our case study to test the conventional wisdom applied in previous studies of this approach \cite{2010PhRvD..81f3512Y} and from CMB weak lensing research. We find that in specific cases this wisdom does not hold. For example, previous studies suggest that using EB will result in the best reconstruction of T to P leakage. However, we find that the TB correlations provide the best reconstruction. In some cases using the Wiener filter, the optimum filter that is used in delensing, resulted in divergence when cleaning. The aforementioned Gaussian filter was found to avoid this divergence. These examples, where the conventional wisdom does not apply in the case of systematics cleaning, show the importance of carrying out this case study.

\par

\par 
\par 

\par
A number of complications exist which will need addressing for this method to be viable that we leave to future work to consider. These complications include the inclusion of foregrounds, and the presence of multiple different systematics. It may be possible to reconstruct and remove these systematics simultaneously. Moreover, is may be possible break the degeneracy between some cosmological signals and systematics using a QE approach in combination with prior knowledge of the scan strategy.
\par 
Despite the additional complications that need to be considered, this detailed study of the QE reconstruction and the improvements made to the iterative cleaning process are an important step towards implementing QEs to reconstruct and remove systematic effects from upcoming CMB surveys. We suggest that this QE technique should be used to compliment traditional systematic correction techniques to diagnose and remove residual contamination in the data not corrected by other methods.
\acknowledgments
JW and NM are supported by Science and Technology Facilities Council (STFC) studentships. DBT acknowledges support from STFC grant ST/T000341/1 and ST/P000649/1. 
AR was supported by the ERC Consolidator Grant {\it CMBSPEC} (No.~725456) as part of the European Union's Horizon 2020 research and innovation program. MLB acknowledges support from STFC grant ST/T007222/1.



\appendix

\section{\label{app:QE appendix}Geometric identity}

The QEs, as derived in Section~\ref{sec:qe}, rely on geometric couplings between the modes of spin-2 and spin-0 fields. Here, we present details of the important geometric terms used in the QE derivations, and derive the identity found in equation~\eqref{eqn:I identity}.
We begin with the integral of the spherical harmonic terms. In general this term can be written in terms of the Wigner-3j symbols,
\begin{eqnarray}
    \int d\uvec{n}\;{}_{s_{1}}Y_{l_{1}m_{1}}(\uvec{n}){}_{s_{2}}Y_{l_{2}m_{2}}(\uvec{n}){}_{s_{3}}Y_{l_{3}m_{3}}(\uvec{n})=&&\,\sqrt{\frac{(2l_{1}+1)(2l_{2}+1)(2l_{3}+1)}{4\pi}}\nonumber\\
    &&\,\times\left(\begin{array}{ccc}
         l_{1}&l_{2}&l_{3}  \\
         m_{1}&m_{2}&m_{3} 
    \end{array}\right)\left(\begin{array}{ccc}
         l_{1}&l_{2}&l_{3}  \\
         -s_{1}&-s_{2}&-s_{3} 
    \end{array}\right)\;.
\end{eqnarray}
Using this identity we see that the geometric term, $\Ipm{Ll_{2}l}{Mm_{2}m}$, that was introduced in equation \eqref{eqn:deltaX} is explicitly written as
\begin{eqnarray}\label{eqn: I definition long}
    \Ipm{Ll_{2}l_{1}}{Mm_{2}m_{1}}=&&
    \int d\hat{n}\,\Yspin{LM}Y_{l_{2}m_{2}}(\uvec{n}){}_{\pm2}Y^*_{l_{1}m_{1}}(\uvec{n})\nonumber\\
    =&&(-1)^{m_{1}}\sqrt{\frac{(2L+1)(2l_{2}+1)(2l_{1}+1)}{4\pi}}\left(\begin{array}{ccc}
         L&l_{2}&l_{1}  \\
          M&m_{2}& -m_{1}
    \end{array}\right)\left(\begin{array}{ccc}
         L&l_{2}&l_{1}  \\
         \mp2&0&\pm2 
    \end{array}\right)\;.
\end{eqnarray}
We can use the coupling parity, $\ell\equiv L+l_{2}+l_{1}$, and the $H^{L}_{l_{2}l_{1}}$ term,
\begin{equation}
    H^{L}_{l_{2}l_{1}}\equiv  \sqrt{\frac{(2L+1)(2l_{2}+1)(2l_{1}+1)}{4\pi}}\left(
    \begin{array}{ccc}
         L & l_{2} & l_{1}  \\
         -2 & 0 & 2\\
    \end{array}
    \right)\;,
\end{equation}
to simplify equation \eqref{eqn: I definition long} giving
\begin{equation}
    \Ipm{Ll_{2}l_{1}}{Mm_{2}m_{1}}=(-1)^{m_{1}}(\pm1)^{\ell}\,H^{L}_{l_{2}l_{1}}\left(\begin{array}{ccc}
         L&l_{2}&l_{1}  \\
          M&m_{2}& -m_{1}\\
    \end{array}\right)\;.
\end{equation}
The identity in equation~\eqref{eqn:I identity} can then be derived starting with
\begin{equation}
    \sum_{m_{2}m_{1}}\Ipm{Ll_{2}l_{1}}{Mm_{2}m_{1}}\,\Ipm{L'l_{2}l_{1}}{M'm_{2}m_{1}} = \sum_{m_{2}m_{1}}(\pm1)^{\ell+\ell'}\,H^{L}_{l_{2}l_{1}}\,H^{L'}_{l_{2}l_{1}}\left(\begin{array}{ccc}
         L&l_{2}&l_{1}  \\
          M&m_{2}& -m_{1}\\
    \end{array}\right)\left(\begin{array}{ccc}
         L'&l_{2}&l_{1}  \\
          M'&m_{2}& -m_{1}\\
    \end{array}\right)\;.\label{eqn:near Ident}
\end{equation}
Carrying out the sum on the right hand side and applying the property of the Wigner-3j symbols,
\begin{equation}
    \sum_{m_{1}m_{2}}(2L+1)\left(\begin{array}{ccc}
         l_{1}&l_{2}&L \\
         m_{1}&m_{2}&M
    \end{array}\right)\left(\begin{array}{ccc}
         l_{1}&l_{2}&L'  \\
         m_{1}&m_{2}&M' 
    \end{array}\right) = \delta_{LL'}\delta_{MM'}\;,
\end{equation}
simplifies the right hand side of \eqref{eqn:near Ident}, giving the identity in \eqref{eqn:I identity},
\begin{equation}
    \sum_{m_{2}m_{1}}\Ipm{Ll_{2}l_{1}}{Mm_{2}m_{1}}\,\Ipm{L'l_{2}l_{1}}{M'm_{2}m_{1}} = \frac{\left(H^{L}_{l_{2}l_{1}}\right)^{2}}{2L+1}\delta_{LL'}\delta_{MM'}\;.
\end{equation}

\bibliographystyle{JHEP.bst}
\bibliography{BlindMapClean}

\end{document}